\documentclass{article}
\usepackage{xparse}
\usepackage{packages} 

\title{APRIL: Auxiliary Physically-Redundant Information in Loss\\\Large A physics-informed framework for parameter estimation with a gravitational-wave case study}

\author[1,2]{Matteo Scialpi\orcidlink{0009-0007-6434-1460}\footnote{Corresponding author: \href{matteo.scialpi@unife.it}{matteo.scialpi@unife.it}}}
\author[3]{Francesco Di Clemente\orcidlink{0000-0002-8257-3819}\footnote{\href{fdicleme@central.uh.edu}{fdicleme@central.uh.edu}}}
\author[4,5]{Leigh Smith\orcidlink{0000-0002-3035-0947}\footnote{\href{leighann.smith@units.it}{leighann.smith@units.it}}}
\author[2,6]{Micha\l~Bejger\orcidlink{0000-0002-4991-8213}\footnote{\href{bejger@fe.infn.it}{bejger@fe.infn.it}}}

\affil[1]{\textit{Dipartimento di Fisica e Scienze della Terra, Università di Ferrara, Via Saragat 1, 44122 Ferrara, Italy}}
\affil[2]{\textit{INFN Sezione di Ferrara, Via Saragat 1, 44122 Ferrara, Italy}}
\affil[3]{\textit{Department of Physics, University of Houston, Houston, TX 77204, USA}}
\affil[4]{\textit{Dipartimento di Fisica, Università di Trieste, Via Valerio 2, 34127 Trieste, Italy}}
\affil[5]{\textit{INFN, Sezione di Trieste, I-34127 Trieste, Italy}}
\affil[6]{\textit{Nicolaus Copernicus Astronomical Center, Polish Academy of Sciences, Bartycka 18, 00-716 Warszawa, Poland}}

\date{\empty}

\begin{document}

\maketitle
\begin{abstract}
    Physics-Informed Neural Networks (PINNs) embed the partial differential equations (PDEs) governing the system under study directly into the training of Neural Networks, ensuring solutions that respect physical laws. While effective for single-system problems, standard PINNs scale poorly to datasets containing many realizations of the same underlying physics with varying parameters. To address this limitation, we present a complementary approach by including auxiliary physically-redundant information in loss (APRIL), i.e. augment the standard supervised output-target loss with auxiliary terms which exploit exact physical redundancy relations among outputs. We mathematically demonstrate that these terms preserve the true physical minimum while reshaping the loss landscape, improving convergence toward physically consistent solutions. As a proof-of-concept, we benchmark APRIL on a fully-connected neural network for gravitational wave (GW) parameter estimation (PE). We use simulated, noise-free compact binary coalescence (CBC) signals, focusing on inspiral-frequency waveforms to recover the chirp mass $\mathcal{M}$, the total mass $M_\mathrm{tot}$, and symmetric mass ratio $\eta$ of the binary. In this controlled setting, we show that APRIL achieves up to an order-of-magnitude improvement in test accuracy, especially for parameters that are otherwise difficult to learn. This method provides a physically consistent training approach for more realistic GW analysis applications.
\end{abstract}

\section{Introduction}
\par Machine learning (ML) methods, and in particular applications based on neural networks (NNs), have become essential tools in engineering and, increasingly, in the physical sciences \cite{MLreview,Cuoco_2021,Cuoco_2025,cuoco2025gravitational}. A NN is a composition of nonlinear operations (neurons) whose parameters are optimized by minimizing a user-defined loss function. In the standard supervised learning setting, network outputs are directly compared to target values, and the parameters are updated to reduce this discrepancy \cite{wetzel2025interpretablemachinelearningphysics}. Such purely data-driven approaches are often extremely effective in industrial applications, where predictive accuracy is the primary goal \cite{fred2024systematicliteraturereviewuse,Haneena_Jasmine_2021}. In physics, however, purely data-driven training could lead to results that are numerically accurate but physically inconsistent. This is because physical quantities are often bound by exact algebraic or phenomenological relations derived from well-established physical laws. In a standard NN, these relationships are not explicitly enforced, and the model can only approximate them numerically, without full interpretability \cite{sun2024reviewmultimodalexplainableartificial}; see a recent focus issue collection of articles addressing ML explainability \cite{IOPFocusEMLIssue2024}.  

To address this issue, physics-informed neural network (PINN) framework was introduced \cite{RAISSI2019686}, in which the underlying physical ``laws'' e.g. governing partial differential equations (PDEs) are taken into account in the loss function. Using automatic differentiation, PDE residuals can be computed from the NN outputs at chosen collocation points, enabling the loss to combine a data term (output-target agreement) and a physics term (PDE residual minimization). PINNs have now found a huge variety of applications, from fluid dynamics to geophysics, see \cite{cuomo2022scientificmachinelearningphysicsinformed} 
for a review.

However, while PINNs are successful in solving inverse and forward problems for a single physical system, they are not well suited for learning from large datasets containing many realizations of the same underlying physics but with varying parameters, see e.g.  \cite{torres2025adaptivephysicsinformedneuralnetworks}. For instance, consider modeling the motion of many pendulums with different lengths and masses. The governing equations (simple harmonic oscillator with gravity as restoring force) are identical for each pendulum, but the parameters (masses and pendula lengths) change from one system to another. A traditional PDE-based PINN would need to be retrained for each realization or handle all parameter sets simultaneously within a single PDE-constrained optimization, which is computationally prohibitive.

To address this limitation, we propose to introduce auxiliary physically-redundant information in losses (APRIL), which retains the PINN philosophy of embedding physics into the training process, but in a way that scales efficiently to datasets with many distinct realizations of the same physical system. APRIL augments the standard output-target loss with auxiliary terms derived from known physical redundancy relations between network outputs. This is a follow up in-depth study to a concept noted by  \cite{diclemente2025explainableautoencoderneutronstar}, where an autoencoder architecture was applied to estimating the parameters of neutron stars' (NS) equation of state (EoS). Here, we expand the mathematical demonstration that these extra terms do not shift the location of the true minimum of the loss function. Instead, they reshape the loss landscape, sharpening the global physical minimum and guiding optimization toward physically valid solutions.

As a proof of concept, we apply APRIL to a problem of gravitational-wave (GW) parameter estimation (PE), see \cite{Cuoco_2025} for a recent review on ML for GW science. GWs from compact binary coalescences (CBCs) \cite{10.1093/acprof:oso/9780198570745.001.0001}, i.e. emitted during last stages of the binary system composed of two compact bodies, are well described by the post-Newtonian (PN) expansion \cite{Cutler_1994}, in which key mass-related quantities (chirp mass $\mathcal{M}$, total mass $M_\mathrm{tot}$ and symmetric mass ratio $\eta$) are linked by exact algebraic relations, since only two masses define the system. These relationships make GW PE an ideal testbed for APRIL: they are simple to encode and physically exact. In our benchmark study, we train a deliberately simple Fully Connected Neural Network (FCNN) on simulated, noise-free GW frequency signals and compare performance with and without APRIL loss terms. This setting isolates the effect of the loss design itself and leads to an improvement in overall PE accuracy by an order of magnitude compared to a purely data-driven loss.

Previous works have already explored the use of PINNs in GW physics \cite{diclemente2025explainableautoencoderneutronstar,Keith_2021,PhysRevD.106.124047,patel2024calculatingquasinormalmodesschwarzschild,PhysRevD.107.064025,sym13112157,Rosofsky_2023,Auddy_2024,10.1093/mnras/stad1810,10.1093/mnras/stad2840,li2023solvingeinsteinequationsusing}. However, to the best of our knowledge, this is the first time that a PINN-inspired network is applied to the direct PE of GW signals, even if in a fully simulated and noise-free setting. Although GW signals are typically hidden in noise, our fully simulated and noise-free setting is sufficient for this methodological proof-of-concept. We regard this study as a proof-of-concept that can serve as the foundation for future algorithms, including more advanced versions currently in preparation, studying how to deal with noise and spectrogram images. For completeness, in Appendix \ref{app:noise} one can find the results for a similar study on the robustness against Gaussian noise added to the frequency signal.

Before going forward with the paper, we note that recent progress in physics-informed ML have already produced several approaches that alleviate the need to retrain a PDE-constrained model for each new realization of the system. For example, meta-learning PINNs \cite{PENWARDEN2023111912} learn initializations that adapt quickly to new PDE instances, neural operators \cite{10.5555/3648699.3648788} such as DeepONets \cite{Lu_2021} and Fourier neural operators  \cite{li2021fourierneuraloperatorparametric} directly learn solution operators across parameter families, and physics-informed neural transformer operators (PINTOs) \cite{boya2025physicsinformedtransformerneuraloperator} incorporate parameterized physics into operator-learning architectures. Rather than learning the governing operator, APRIL reshapes the loss landscape so that standard supervised NN models learn physically consistent mappings more effectively. APRIL provides a lightweight way to embed exact algebraic relations and physically redundant constraints directly into the training loss, without requiring explicit PDE residual computation, collocation points, or the optimization overhead of a strong-form PINN. In this sense, APRIL is complementary to the above families of methods. For problems like GW PE, where the inputs are not fields governed by a PDE, but derived features (e.g. frequency–time tracks), and where only certain combinations of outputs obey known analytic relations, APRIL supplies an efficient physics-based regularization that these operator-learning methods do not directly target.

The paper is organized as follows. In Sec. \ref{sec:NN_and_loss}, we present the theoretical framework of APRIL and analyze its effect on the optimization landscape. Section~\ref{sec:GW benchmark} describes the GW case study, including the physical context (Sec. \ref{sec:physical_context}), methodology (Sec. \ref{sec:methodology}), and results (Sec. \ref{sec:discussion}). We summarize our findings and discuss future extensions in Sec. \ref{sec:conclusions}. The paper is supplemented by appendices \ref{app:noise}, \ref{app:03_BPL2P} and \ref{app:extended_results}, describing tests with noisy data, the input data based on realistic mass distribution and extended results covering all loss terms, respectively.

All the codes related to the benchmark study are developed using the PyTorch library; see \cite{APRIL-implementation} for the implementation and saved models.

\section{Loss function in a neural network}
\label{sec:NN_and_loss}

We will introduce here a general definition of a loss function, to subsequently discuss improvements based on auxiliary physically-informed additions. 

Consider a dataset $\left\{\textbf{x}_i\right\}_{i=1}^D$, where $\textbf{x}_i\in\mathbb{R}^\mathrm{m}$ could be experimental or simulated data, and $i$ is the index that defines the particular realization of the system. In addition, a set of features $\left\{\hat{\textbf{y}}_i\right\}_{i=1}^D$, with $\hat{\textbf{y}}_i\in\mathbb{R}^\mathrm{n}$ related to $\textbf{x}_i$ thanks to a (not directly accessible) map $\mathcal{F}$ is 
\begin{equation}
    \hat{\textbf{y}}_i=\mathcal{F}(\textbf{x}_i)\,.
\end{equation}
In this work, we use the term \textit{ground truth} to denote the actual physical quantities, defined independently of the neural network. When the same quantities are employed within the algorithm as reference values for the outputs to approximate, we designate them as \textit{targets}.
\par A NN algorithm aims to represent $\mathcal{F}$ thanks to a non-linear combination of different functions. For example, the simplest case of a feed-forward deep NN composed of two hidden layers can be written as
\begin{equation}
\textbf{y}_{\theta,i}=W_3\sigma_2\left(W_2\sigma_1\left(W_1\textbf{x}_i+b_1\right)+b_2\right)+b_3\,,
    \label{eq:ffnn}
\end{equation}
where $W$ are the NN weights, $b$ the biases, $\sigma$ the chosen activation functions, $\theta=(W_1,W_2,W_3,b_1,b_2,b_3)$ the NN parameters, and $\textbf{y}_{\theta,i}$ the NN representation for the target features $\hat{\textbf{y}}_i$ with that particular choice for parameters $\theta$.

NN's well-structured architecture depends on the desired task to be accomplished and on the kind of data that one has to handle \cite{HORNIK1989359,726791,rumelhart1986learning,rezende2016variationalinferencenormalizingflows,vaswani2023attentionneed,Neal1996}. While these architectures are usually too complicated to be expressed with a simple equation such as \cref{eq:ffnn}, in general the NN is a transformation  
\begin{equation}
    \textbf{y}_{\theta,i}=\textrm{NN}_\theta(\textbf{x}_i)\,.
\end{equation}
The ideal case for the NN will be to perfectly represent the map $\mathcal{F}$, leading to
\begin{equation}
    \hat{\textbf{y}}_i=\textbf{y}_{\theta^*,i}=\textrm{NN}_{\theta^*}(\textbf{x}_i)=\mathcal{F}(\textbf{x}_i)\,,
    \label{eq:NN_action}
\end{equation}
where $\theta^*$ is the optimal choice for $\theta$. This optimization problem for $\theta$ can be accomplished thanks to the minimization of the loss function $\mathcal{L}$, usually expressed in terms of a distance. In many applications, loss is defined as the Mean Square Error (MSE) between the predicted $\textbf{y}_{\theta,i}$ and the target value $\hat{\textbf{y}}_i$ \cite{rumelhart1986learning}:
\begin{equation}
    \mathcal{L}_\mathrm{t}(\theta)=\mathrm{MSE}(\textbf{y}_{\theta,i},\hat{\textbf{y}}_i)=\frac{1}{B}\sum_{i=1}^B\left(\textbf{y}_{\theta,i}-\hat{\textbf{y}}_i\right)^2\,,
    \label{eq:data_loss}
\end{equation}
where $B$ is the batch size; MSE here is proportional to the L2 (Euclidean) distance. The training implementation permits, thanks to an optimizer algorithm, to find the value for $\theta^*$ finding the minimum for $\mathcal{L}$, which by construction will be
\begin{equation}
    \min_\theta\mathcal{L}_\mathrm{t}(\theta)=\mathcal{L}_\mathrm{t}(\theta^*)=0\,.
\end{equation}
\par In summary, the goal of the NN training process is to identify the optimal parameters $\theta^*$ that allow the model to faithfully approximate the {\it a priori} unknown map $\mathcal{F}$ from the data alone, providing a purely data-driven baseline before incorporating any additional physical constraints.

\subsection{Auxiliary physics-informed loss}
\label{sec:Auxiliary Physics-Informed Loss}

Neural networks in engineering typically learn mappings between inputs and outputs without considering the underlying physical relationships that govern these quantities. While standard loss functions evaluate each output independently against its target, physical systems impose cross-constraints between variables through conservation laws, symmetries, and governing equations. We propose that incorporating this physical knowledge directly into the loss function — through what we call auxiliary physically-redundant information - fundamentally improves the optimization landscape and accelerates convergence toward physically consistent solutions.

Let $\hat{\mathbf{y}}_i = (\hat{y}_{1,i}, \hat{y}_{2,i}, \ldots, \hat{y}_{n,i})^T$ denote the ground truth features and $\mathbf{y}_{\theta,i} = (y_{\theta,1,i}, y_{\theta,2,i}, \ldots, y_{\theta,n,i})^T$ represent the network outputs for input $i$ with parameters $\theta$. Physical systems often impose well-defined relationships among these quantities. Consider a known physical constraint linking the ground truth features:
\begin{equation}
    \hat{y}_{1,i}=g(\hat{y}_{2,i},\hat{y}_{3,i},\dots,\hat{y}_{n,i})\,.
    \label{eq:hat_y}
\end{equation}

This relationship might encode energy conservation, momentum balance, thermodynamic equations of state, or other domain-specific laws. Rather than hoping the network discovers these relationships from data alone, we directly embed them into the training process.

Alongside the standard data-driven loss $\mathcal{L}_\mathrm{t}(\theta)$, we introduce auxiliary physics-informed loss terms:
\begin{equation}
    \mathcal{L}_\mathrm{APRIL}(\theta)=\mathrm{MSE}\bigl(g(y_{\theta,2,i},y_{\theta,3,i},\dots,y_{\theta,n,i}),y_{\theta,1,i}\bigr)\,,
    \label{eq:L_physics}
\end{equation}
which measures how well the network outputs satisfy the known physical relationships. The total loss function becomes:
\begin{equation}
    \mathcal{L}_\mathrm{total}(\theta)=\mathcal{L}_\mathrm{t}(\theta)+\lambda\mathcal{L}_\mathrm{APRIL}(\theta)\,,
\end{equation}
where the hyperparameter $\lambda > 0$ balances data fidelity against physical consistency.

A crucial property emerges when analyzing the global minimum. For the mathematical analysis, consider first the idealized case where the ground truth perfectly satisfies physical laws. If the network achieves perfect reconstruction $y_{\theta^*,j,i} = \hat{y}_{j,i}$ for all components $j$ and samples $i$, then:
\begin{equation}
    \mathcal{L}_\mathrm{APRIL}(\theta^*) = \mathrm{MSE}\bigl(g(\hat{y}_{2,i},\hat{y}_{3,i},\dots,\hat{y}_{n,i}),\hat{y}_{1,i}\bigr) = 0\,.
\end{equation}

Hence, APRIL preserves the optimal solution structure while fundamentally altering the learning dynamics. The key insight—which holds regardless of noise—is that the physics-informed terms reshape the loss landscape to favor physically consistent parameter configurations throughout training, enforcing physical consistency. Even when real data contains noise or model approximations, causing $\mathcal{L}_\mathrm{APRIL}(\theta^*) > 0$, the network learns representations that encode physical relationships more robustly than standard approaches.

In noisy scenarios, this creates a beneficial trade-off because while the global minimum of $\mathcal{L}_\mathrm{total}$ may no longer coincide with perfect data reconstruction, the optimization lead to solutions that better capture the underlying physics. The network learns not only the input-output mapping but also how the output components correlate with each other through physical laws. This dual learning, where the input-output relationship is simultaneously constrained by inter-output physical dependencies, produces a more robust and physically grounded model. Standard networks might achieve similar training loss by independently fitting each output, but APRIL-trained networks exploit the system dynamics information, leading to superior performance and interpretability.

Real physical systems typically obey multiple constraints simultaneously. We generalize our approach to handle $K$ distinct physical relationships:
\begin{equation}
    \mathcal{L}_\mathrm{APRIL}(\theta) = \sum_{k=1}^{K} \lambda_k \mathrm{MSE}\bigl(g_k(\mathbf{y}_{\theta,i}), 0\bigr)\,,
    \label{Lapril}
\end{equation}
where each function $g_k$ encodes a different physical principle (formulated to equal zero when satisfied), and individual weights $\lambda_k$ allow fine-tuned control over the relative importance of each constraint.

\subsection{Effects of physics-informed loss terms on the global minimum}
\label{sec:global_minimum_effects}

To understand how APRIL transforms the optimization problem, we analyze the local geometry of the loss landscape around the global minimum $\theta^*$. Assuming perfect data fitting at this point, we have:
\begin{equation}
    \mathcal{L}_\mathrm{t}(\theta^*)=0\,,\qquad\mathcal{L}_\mathrm{APRIL}(\theta^*)=0\,.
\end{equation}

Since $\theta^*$ minimizes both loss components, their gradients must vanish:
\begin{equation}
    \nabla\mathcal{L}_\mathrm{t}(\theta^*)=\mathbf{0}\,,\qquad\nabla\mathcal{L}_\mathrm{APRIL}(\theta^*)=\mathbf{0}\,.
\end{equation}

Therefore, the gradient of the total loss also vanishes:
\begin{equation}
    \nabla\mathcal{L}_\mathrm{total}(\theta^*) = \nabla\mathcal{L}_\mathrm{t}(\theta^*) + \lambda\nabla\mathcal{L}_\mathrm{APRIL}(\theta^*) = \mathbf{0} + \lambda\cdot\mathbf{0} = \mathbf{0}\,.
\end{equation}

Using a second-order Taylor expansion around $\theta^*$, we can approximate the total loss for nearby parameters:
\begin{equation}
    \mathcal{L}_{\text{total}}(\theta)\approx\mathcal{L}_{\text{total}}(\theta^*)+\nabla\mathcal{L}_\mathrm{total}(\theta^*)^T\Delta\theta+\frac{1}{2}\Delta\theta^TH_{\mathcal{L}_{\text{total}}}\Delta\theta\,,
\end{equation}
where $\Delta\theta=\theta-\theta^*$ represents the parameter displacement from the minimum, and $H_{\mathcal{L}_{\text{total}}}$ is the Hessian matrix of second derivatives evaluated at $\theta^*$.

Given that both the loss value and gradient vanish at $\theta^*$, this simplifies to:
\begin{equation}
    \mathcal{L}_{\text{total}}(\theta)\approx\frac{1}{2}\Delta\theta^TH_{\mathcal{L}_{\text{total}}}\Delta\theta\,.
\end{equation}

By the linearity of differentiation, the Hessian of the combined loss equals the sum of the individual Hessians:
\begin{equation}
    H_{\mathcal{L}_{\text{total}}}(\theta^*)=H_{\mathcal{L}_\mathrm{t}}(\theta^*) + \lambda H_{\mathcal{L}_\mathrm{APRIL}}(\theta^*)\,,
\end{equation}
where $H_{\mathcal{L}_\mathrm{t}}$ and $H_{\mathcal{L}_\mathrm{APRIL}}$ are the Hessians of $\mathcal{L}_\mathrm{t}(\theta)$ and $\mathcal{L}_\mathrm{APRIL}(\theta)$ at $\theta^*$, respectively. Since both $\mathcal{L}_\mathrm{t}(\theta)$ and $\mathcal{L}_\mathrm{APRIL}(\theta)$ are constructed from squared-error terms, their Hessians are positive semi-definite:
\begin{equation}
    H_{\mathcal{L}_\mathrm{t}}\succeq0\,,\qquad H_{\mathcal{L}_\mathrm{APRIL}}\succeq0\,.
\end{equation}

This property guarantees that the eigenvalues of both matrices are non-negative, ensuring the minimum is not a saddle point. For any parameter direction $\Delta\theta$, the quadratic form representing curvature in that direction satisfies:
\begin{equation}
    \Delta\theta^TH_{\mathcal{L}_{\text{total}}}\Delta\theta=\Delta\theta^TH_{\mathcal{L}_\mathrm{t}}\Delta\theta+\lambda\Delta\theta^TH_{\mathcal{L}_\mathrm{APRIL}}\Delta\theta\,.
\end{equation}

Since both terms are non-negative, we obtain the inequality:
\begin{equation}
    \Delta\theta^TH_{\mathcal{L}_{\text{total}}}\Delta\theta\geq\Delta\theta^TH_{\mathcal{L}_\mathrm{t}}\Delta\theta\,,
\end{equation}
demonstrating that $\mathcal{L}_{\text{total}}(\theta)$ exhibits at least as much curvature as $\mathcal{L}_\mathrm{t}(\theta)$ in every direction. Then, we can express these Hessians through their spectral decompositions:
\begin{equation}
    H_{\mathcal{L}_\mathrm{t}} = V_t \Lambda_t V_t^T\,, \qquad H_{\mathcal{L}_\mathrm{APRIL}} = V_p \Lambda_p V_p^T\,,
\end{equation}
where $V_t$ and $V_p$ are orthogonal matrices whose columns are the eigenvectors, while $\Lambda_t = \text{diag}(\lambda_{t,1}, \lambda_{t,2}, \ldots)$ and $\Lambda_p = \text{diag}(\lambda_{p,1}, \lambda_{p,2}, \ldots)$ are diagonal matrices containing the non-negative eigenvalues. In order to quantify the sensitivity of the algorithm's output to small changes in its input, or specifically in our case, to study the convergence rates for gradient-based optimization, we employ the condition number, defined as:
\begin{equation}
    \kappa(H_{\mathcal{L}_\text{total}}) = \frac{\lambda_\text{max}(H_{\mathcal{L}_\text{total}})}{\lambda_\text{min}(H_{\mathcal{L}_\text{total}})}\,.
\end{equation} 
Consider the critical case where $H_{\mathcal{L}_\mathrm{t}}$ has very small eigenvalues along certain directions—these correspond to nearly flat regions in the loss landscape where different parameter configurations yield similar network outputs. Without physical constraints, gradient descent can converge to any of these degenerate minima. APRIL breaks this degeneracy by adding curvature specifically along physically inconsistent directions. If $H_{\mathcal{L}_\mathrm{APRIL}}$ contributes positive eigenvalues along these previously flat directions, then:
\begin{equation}
    \lambda_\text{min}(H_{\mathcal{L}_\text{total}}) = \lambda_\text{min}(H_{\mathcal{L}_\mathrm{t}}) + \lambda \cdot \lambda_\text{min}(H_{\mathcal{L}_\mathrm{APRIL}}) > \lambda_\text{min}(H_{\mathcal{L}_\mathrm{t}})\,.
\end{equation}

This selective injection of curvature creates a fundamental asymmetry: parameter configurations that satisfy physical constraints remain at local minima, while physically inconsistent solutions are lifted to higher loss values. Among the originally degenerate minima of $\mathcal{L}_\mathrm{t}$, APRIL preserves only those that respect the physical laws, effectively using physics as a selection principle.

\subsection{Regularization effect}
\label{sec:regularization}

Enhanced curvature due to physics-informed terms creates a steeper, more well-defined gradient field near $\theta^*$. This modification of the loss landscape provides two distinct advantages.

Firstly, it eliminates spurious ``flat'' regions. Parameter space often contains manifolds where different weights produce nearly identical outputs, creating valleys with minimal curvature. APRIL lifts these degeneracies by penalizing physically inconsistent parameter combinations, even if they produce similar network outputs.

Secondly, it provides physics-informed regularization. Unlike $L_1$ or $L_2$ regularization that blindly penalize parameter magnitudes, APRIL specifically targets solutions that violate domain knowledge. The network remains free to use large parameters when necessary for accuracy, but only along trajectories that respect physical constraints.

In practical applications, perfect data fitting rarely occurs. Measurement noise, model limitations, and numerical approximations typically lead to a residual error $\mathcal{L}_\mathrm{t}(\theta^*) \approx \epsilon$ for some small $\epsilon > 0$. Under these conditions, the Taylor analysis remains approximately valid, and the physics-informed terms continue to guide optimization toward physically consistent solutions, potentially achieving lower values of $\mathcal{L}_\mathrm{APRIL}$ even at the expense of slightly higher data-fitting error.

Selecting appropriate values for the weighting parameters $\lambda_k$ in Sec.~\ref{Lapril} requires balancing competing objectives. If too small, the physical constraints provide negligible guidance, if too large, they dominate the data-fitting objective, potentially preventing the network from learning important patterns not captured by simplified physical models.

\section{Gravitational-wave parameter estimation losses benchmark}
\label{sec:GW benchmark}
We will apply here the methodology described in the previous section to a real science case: the solution for the GW CBC signal inverse problem. The aim is to infer the value of the mass-related parameters $M_\textrm{tot}$, $\eta$, and $\mathcal{M}$ (defined below) from GW frequency data with a ML solution. 

During the training stage we will perform the PE using different auxiliary loss terms. Successfully trained models will be frozen and tested on a common test dataset, independent from the training ones. By evaluating the accuracy thanks to a common metric, we will constitute a benchmark study for the training with different loss terms.

\subsection{Gravitational-wave signal waveform}
\label{sec:physical_context}

GWs are space-time perturbations, a class of wave-like solutions in the  General Relativity (GR), which were found formulation of the theory \cite{Einstein1915,Einstein1916,Einstein1918}. 
Experimental, computational, and theoretical efforts are together at work to explore this new way of observing the Universe. So far, the only astrophysical sources detected with the currently active ground-based instruments - Advanced Laser Interferometer Gravitational Observatory (LIGO) \cite{LIGO2015}, Advanced Virgo \cite{Virgo2015} KAmioka GRavitational-wave Antenna (KAGRA) \cite{KAGRA2021}, and GEO600 \cite{Dooley_2015} operating by the LIGO-Virgo-KAGRA (LVK) collaboration - are the CBCs, where the compact components - NSs or BHs - spiral toward each other until they merge (compactness in this context means that the size is close to or equal to the Schwarzschild radius \cite{Schwarzschild1916}). 

GWs carry away the kinetic energy of the binary which means that the orbital velocity increases, and the separation between components decreases. The dominant GW frequency is twice the orbital frequency until the quasi-circular inspiral becomes unstable, and the components merge into one object. In the following we will focus on the inspiral part of the GW waveform.  

Denoting $m_1$ and $m_2$ the component masses, one can define 
\begin{equation}
    \mathcal{M}=\frac{\left(m_\textrm{1}m_\textrm{2}\right)^{3/5}}{\left(m_\textrm{1}+m_\textrm{2}\right)^{1/5}}\,,\qquad M_\textrm{tot}=m_\textrm{1}+m_\textrm{2}\,,\qquad\eta=\frac{m_\textrm{1}m_\textrm{2}}{(m_\textrm{1}+m_\textrm{2})^2}\,,
    \label{eq:mass_related}
\end{equation}
where $\mathcal{M}$ is the chirp mass, $M_\textrm{tot}$ the total mass, and $\eta$ the symmetric mass ratio. Note that $0<\eta\leq0.25$ is the range of physically meaningful values of this parameter, with $\eta = 0.25$ is corresponding to the equal-mass binary, $m_1=m_2$. Since the three quantities depend only on the two component masses, one can easily see the following relation among the three:
\begin{equation}
    \mathcal{M}=M_\textrm{tot}\eta^{3/5}\,.
    \label{eq:mass_bound}
\end{equation}

On one hand GW theory was discovered as a consequence of the linearization of GR, on the other hand the full dynamics of the inspiral orbit (and so the emitted GWs) needs to be described by Einstein's field equations. Unfortunately, the latter are too difficult to solve analytically. An approach for studying GWs in detail is the Post-Newtonian (PN) framework, where expansion GR terms are added to the main Newtonian dynamics of the studied system. Adding more terms, the number of parameters increases and the physics becomes more similar to GR dynamics \cite{Blanchet_2002}. Concerning CBCs, we can express the frequency time derivative of the emitted GWs thanks to this PN formalism, which, at 1.5 order, includes precisely the mass-related quantities as in the following differential equation \cite{Cutler_1994}:
\begin{equation}
    \frac{df}{dt}=\frac{96}{5}\pi^{8/3}\left(\frac{GM_\odot}{c^3}\frac{\mathcal{M}}{M_\odot}\right)^{5/3}f^{11/3}\left[1-\left(\frac{743}{336}+\frac{11}{4}\eta\right)\varepsilon+4\pi\varepsilon^{3/2}\right]\,,
    \label{eq:1.5PN_f}
\end{equation}
where 
\begin{equation}
    \varepsilon=\left(\frac{GM_\odot}{c^3}\frac{M_\textrm{tot}}{M_\odot}\pi f\right)^{2/3}
    \label{eq:1.5PN_eps}
\end{equation}
is the expansion parameter, proportional to the orbital velocity of the system, with $G$ the Newton's universal gravity constant, $M_\odot$ the mass of the Sun, and $c$ the speed of light. The term before the squared parentheses is the Newtonian formalism for the inspiral, while the square bracket contains the PN correction terms that describe the phenomenon with more and more accuracy. Notice that the relevant physical quantities, besides the frequency $f$ itself, are only the mass related ones. Adding more GR terms, i.e. going to higher PN terms, will add different quantities to the game, like spins and tidal deformabilities. 

Formally, for a two body system, one may define an innermost-stable circular orbit (ISCO); for smaller separations, the system becomes unstable. The ISCO frequency, at the Newtonian level, is given by
\begin{equation}
    f_\textrm{ISCO}=\left(2^{3/2}\pi\,\frac{GM_\odot}{c^3}\frac{M_\textrm{tot}}{M_\odot}\right)^{-1}\,,
    \label{eq:f0}
\end{equation}
corresponding to the \cref{eq:1.5PN_f} solution at the change of phase between inspiral motion and free-fall. For more details and textbook introduction, see e.g.  \cite{10.1093/acprof:oso/9780198570745.001.0001,2016}.

\subsection{Methodology}
\label{sec:methodology}
We will now describe the simulated datasets, the network used for the benchmark study and the benchmark itself. The dataset (Sec.~\ref{sec:generation_procedure}) and the algorithm (Sec.~\ref{sec:Architecture}) are purposefully simple in order to focus our study on the impact of the loss components. For this reason we choose to work with noiseless time-frequency signals, which anyway are not the real-world realizations of the phenomenon. Our choice is an approximation of the latter, done specifically for this proof-of-concept.

\subsubsection{Dataset}
\label{sec:generation_procedure}

\begin{figure}[t!]
    \centering
    \includegraphics[width=0.69\textwidth]{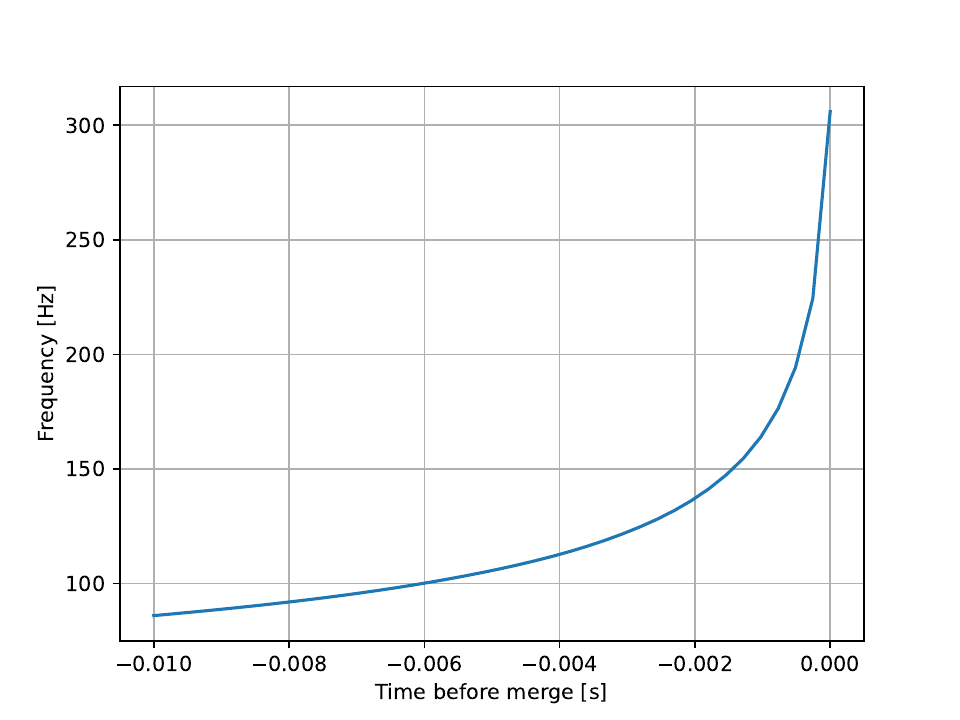}
    \caption{\emph{An example of $f(t_k)$ from the 1.5PN CBC GW event, corresponding to $m_1\simeq59.5\,M_\odot$, $m_2\simeq15.3\,M_\odot$ ($\mathcal{M}\simeq25.2$ M$_\odot$, $M_\textrm{tot}\simeq74.8$ M$_\odot$, and $\eta\simeq0.163$).}}
    \label{fig:dataset}
\end{figure}

\paragraph{Training and validation dataset.} The dataset is created by selecting $D$ samples of  $\left(M_\textrm{tot},\eta\right)$ points from  uniform distributions, $\mathcal{U}\left(\left[40,100\right]\text{ M}_\odot\right)$ for $M_\textrm{tot}$, and $\mathcal{U}\left(\left[0.1,0.25\right]\right)$ for $\eta$, and creating values of $\mathcal{M}$ according to \cref{eq:mass_bound}. With these quantities and \cref{eq:1.5PN_f}, we then compute $\left.\tfrac{df}{dt}\right|_k$ values in the discrete time range $t_k\in\left[-0.01,0\right]$ s, with $t=0$ the time at the quasi-circular inspiral ends, with  $k=1,\dots,K$ corresponding to time steps. For the GW data sampled at a sampling rate $sr=4096$ Hz, $K=40$. Specifically, we store the $\left.\tfrac{df}{dt}\right|_k$ values (the Newtonian term) and the PN corrections separately, 
\begin{align}
    \label{eq:dfdt_newt_separation}
    \left.\frac{df}{dt}\right|_{\mathrm{Newt},\,k}&=\frac{96}{5}\pi^{8/3}\left(\frac{GM_\odot}{c^3}\frac{\mathcal{M}}{M_\odot}\right)^{5/3}f_k^{11/3}\,\\
    \left.\frac{df}{dt}\right|_{\mathrm{corr},\,k}&=1-\left(\frac{743}{336}+\frac{11}{4}\eta\right)\varepsilon_k+4\pi\varepsilon_k^{3/2}\,,
    \label{eq:dfdt_pn_separation}
\end{align}
with $\varepsilon_k$ as in \cref{eq:1.5PN_eps}; note that $\left.\tfrac{df}{dt}\right|_{\mathrm{Newt},\,k}$ depends only on $\mathcal{M}$, while $\left.\tfrac{df}{dt}\right|_{\mathrm{corr},\,k}$ on $M_\textrm{tot}$ and $\eta$. We will study the importance of both terms on loss functions in Sec.~\ref{sec:loss}. 

Integrating then the full form of $\left.\tfrac{df}{dt}\right|_k$ in time thanks to the 4th order Runge-Kutta method \cite{Runge1895,Kutta}, we recover the GW frequency $f_k$. To be more consistent with the physics of the merge, we integrate the differential equation in the backward direction with respect to time, starting from the approximation of the $f_\mathrm{ISCO}$ from \cref{eq:f0}.

Summarizing, an input dataset is composed of $D$ frequency arrays $\left\{f_k\right\}_{k=1}^{K}=f(t_k)$, as shown in Fig.~\ref{fig:dataset}. Alongside the input dataset, $\left\{\hat{\mathcal{M}},\hat{M}_\textrm{tot},\hat{\eta},\left.\frac{df}{dt}\right|^\wedge_{\mathrm{Newt},\,k},\left.\frac{df}{dt}\right|^\wedge_{\mathrm{corr},\,k}\right\}$ values are stored as target.

\begin{figure}[t!]
    \centering
    \subfigure[]{\includegraphics[width=0.32\linewidth]{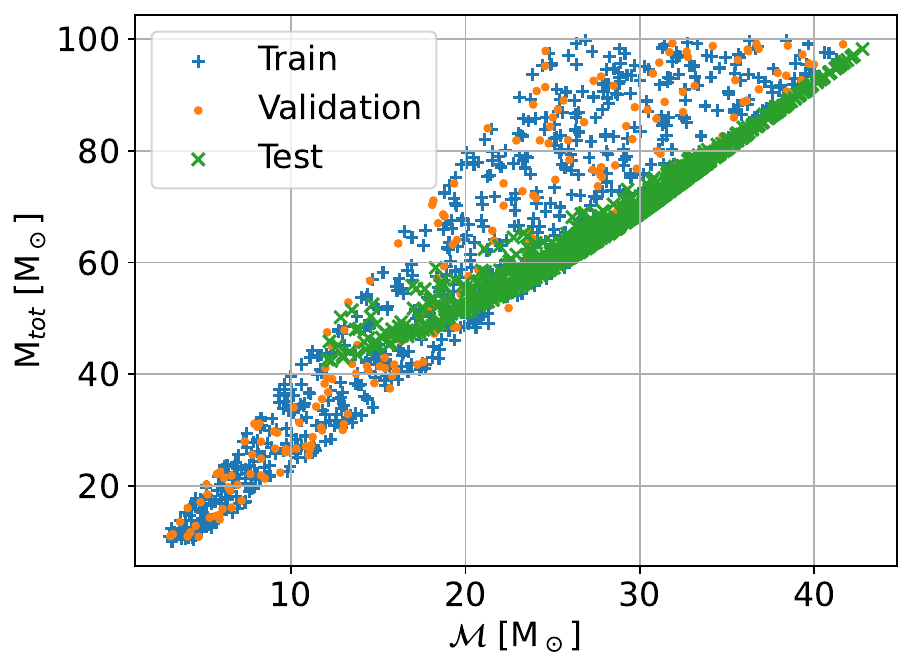}}
    \subfigure[]{\includegraphics[width=0.32\linewidth]{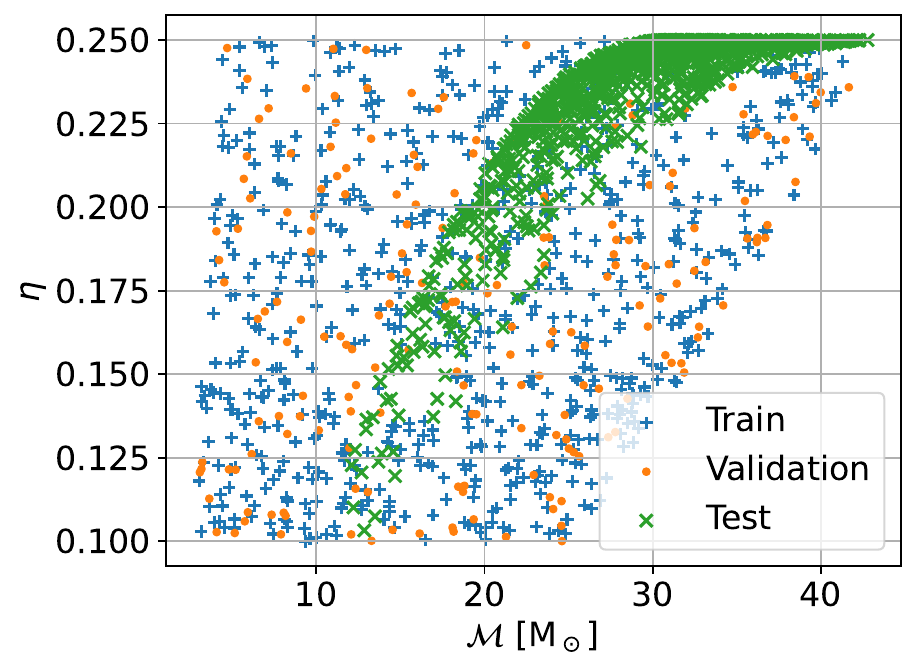}}
    \subfigure[]{\includegraphics[width=0.32\linewidth]{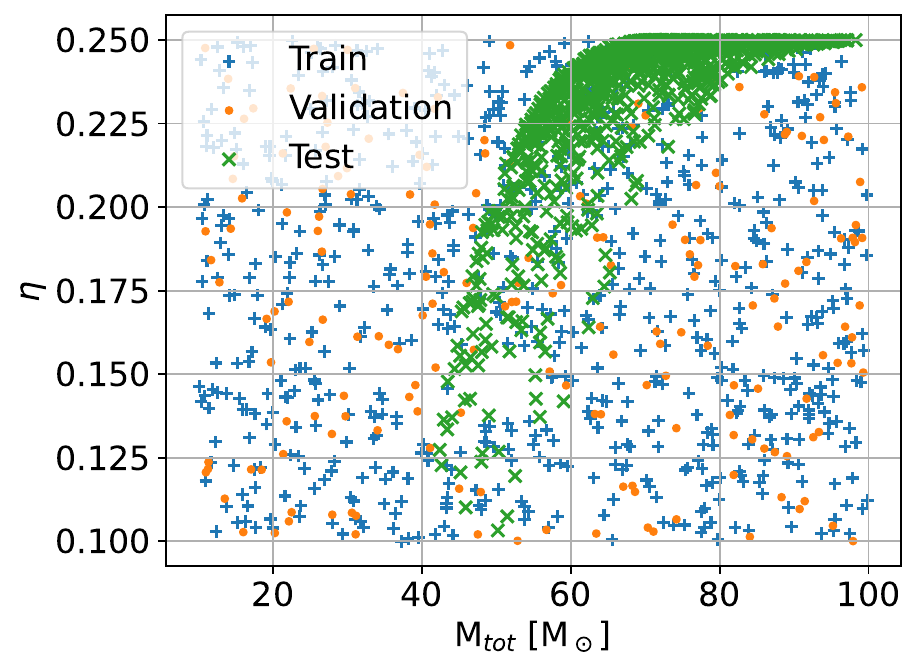}}
    \caption{\emph{Training, validation and test datasets for $\mathcal{M}$, $M_\mathrm{tot}$ and $\eta$. Training and validation datasets are generated by sampling $M_\mathrm{tot}$ and $\eta$ from a uniform distribution. The test dataset is obtained by sampling $m_1$ and $q$ from the mass distribution inferred by LVK collaboration from the GWTC-4 catalog \cite{theligoscientificcollaboration2025gwtc40updatinggravitationalwavetransient,theligoscientificcollaboration2025gwtc40populationpropertiesmerging}, as described in Sec. \ref{sec:test_dataset}.}}
    \label{fig:trainval_test}
\end{figure}

During this analysis we study the performances of the algorithms varying the training dataset size. We build three different datasets with $D=\left\{1,2,5\right\}\times10^3$ instances. They are created with the same procedure and the same random generation seed. From the $D$ signals, the 70\% is the training dataset, while the 20\% is the validation set and the 10\% as test dataset. The latter are only generated from the same simulation process, but there was no interchange between the two after their creation. Examples of the training and validation datasets for $D=10^3$ are shown in \cref{fig:trainval_test}. The built-in test dataset is only used after the last step of the training to roughly compare models among different repetitions. Since the benchmark study is performed on a common test dataset, this built-in dataset is not relevant for the interest of this paper and it is not shown.

\paragraph{Common test dataset.}
\label{sec:test_dataset}
Trained models, described from Sec.~\ref{sec:loss}, need to be tested on the same test dataset to provide a fair benchmark study. We generate a dataset as similar as possible to a set of real GW events: we sample $T=10^3$ data points from the $m_1$ and $q$ ($=m_2/m_1$) broken power-law with 2 Peaks (BPL2P) distribution, approximating the BH population inferred from the first part of the fourth observing run of the LVK  \cite{theligoscientificcollaboration2025gwtc40populationpropertiesmerging}; see also the accompanying series of publications describing the GWTC-4 catalog   \cite{theligoscientificcollaboration2025gwtc40updatinggravitationalwavetransient}. The details are listed in Appendix \ref{app:03_BPL2P}.

Our sampling is performed with the rejection sampling \cite{vonNeumann1951,RobertCasella2004}, a technique for generating samples from a distribution that is difficult to sample directly. We sample $m_1$ and $q=\tfrac{m_2}{m_1}$ from uniform distributions $\mathcal{U}\left(\left[35.5,50.0\right]\text{ M}_\odot\right)$ and $q\in[0.127,1.000]$. Thanks to the sampled values, $\left\{\mathcal{M},M_\mathrm{tot},\eta\right\}$ are computed and so the other quantities. The test dataset can be seen in \cref{fig:trainval_test}.

The domains are so strict in order to recover precisely the training dataset domains for $\left\{\mathcal{M},M_\mathrm{tot},\eta\right\}$. Implications of this and possible strategies to extend the mass domain will be discussed in Sec. \ref{sec:conclusions}.

\subsubsection{Architecture}
\label{sec:Architecture}
\begin{figure}[t!]
    \centering
    \includegraphics[width=1.\textwidth]{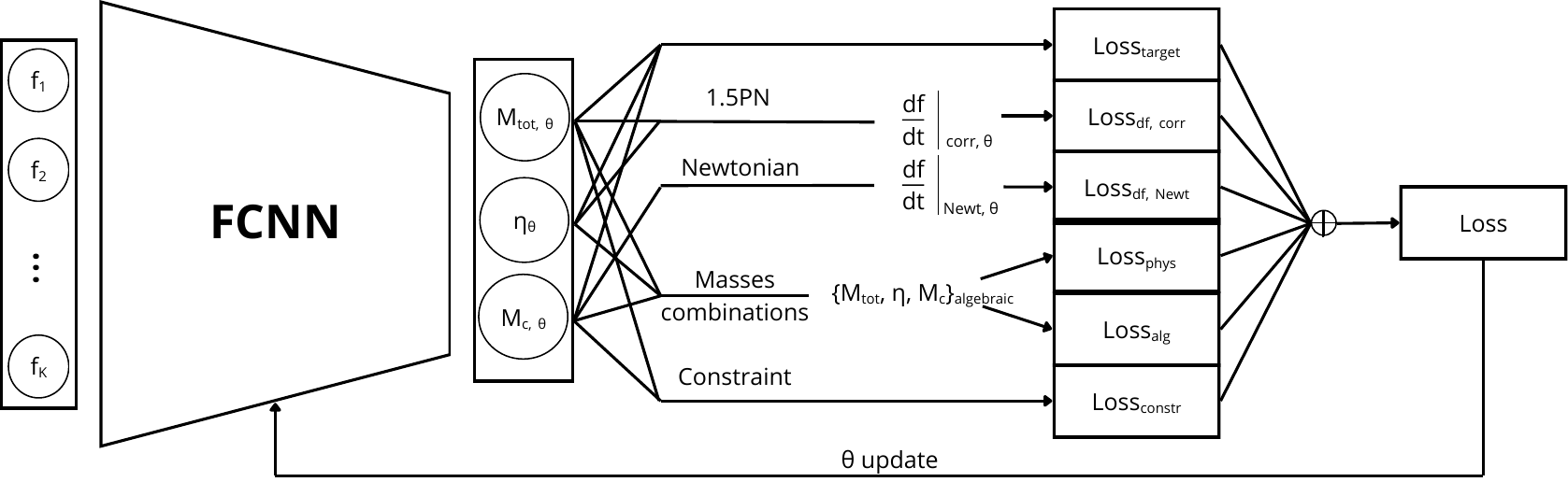}
    \caption{\emph{Algorithm training flow. The frequency array $\left\{f_k\right\}_k^K$ is given as input to the FCNN architecture, to give $\left\{\mathcal{M},M_\mathrm{tot},\eta\right\}_\theta$ outputs. The outputs are then combined in different algebraic quantities that will be substituted in the different loss terms. The total loss is then computed as the sum of all terms. Its value and the gradient meta-data permits to update the NN parameter $\theta$ for a new epoch.}}
    \label{fig:flow}
\end{figure}
The architecture of the algorithm is shown in Fig.~\ref{fig:flow}. 
We have performed a limited grid search to identify a set of suitable values for hyperparameters, adequate to the main goal of this work, i.e. testing the APRIL approach. The impact of batch size and training data size are discussed in detail in the following sections.

The input frequency array $\left\{f_k\right\}_{k=1}^K$ is given as input to a FCNN, with 6 hidden layers. The first 7 layers consists of $2^{12-r}$ neurons each, where $r=1,\dots,7$ is the layer number. This results in the first layer having 2048 neurons, while the input has only size $K=40$. This was chosen in order to have a sufficiently large FCNN, capable of approximating complex functions, and to store as much information as possible in the NN, without down-sampling too early.

The output layer has only three neurons, giving the output values for $\left\{\mathcal{M},M_\mathrm{tot},\eta\right\}_\theta$. The three outputs of the FCNN are normalized $\in\left[0,1\right]$, since a sigmoid activation function is applied as the last NN layer \cite{Verhulst1838,rumelhart1986learning}. Their values are then de-normalized to the training dataset range of values. The three outputs are then combined in different algebraic quantities in order to build the physics-informed loss function $\mathcal{L}_\mathrm{APRIL}$, the details of which are explained in Sec.~\ref{sec:loss}.

During the training, we use the \verb|AdamW| optimizer \cite{Kingma2015,Loshchilov2019}. The \verb|ReduceLROnPlateau| scheduler \cite{Bottou2012,PyTorchReduceLROnPlateau} is also implemented, to reduce the learning rate $lr$ by a factor 10 when the validation loss hadn't decreased in 200 epochs. The initial value is set to $lr=10^{-3}$, while an early stopping is implemented when $lr\leq10^{-8}$. Despite the use of GPUs, we constrain the hardware with deterministic algorithms for reproducibility, both using the \verb|use_deterministic_algorithms| function of PyTorch \cite{PyTorchdeterministic} and fixing a seed for the dataset generation and the training path.

\subsubsection{Components of the loss function}
\label{sec:loss}

The final goal of the training algorithm is the minimization of the loss $\mathcal{L}$ with respect to the NN parameters $\theta$, as we described in Sec.~\ref{sec:NN_and_loss}. As we can see in Fig.~\ref{fig:flow}, the NN outputs $\left\{\mathcal{M},M_\mathrm{tot},\eta\right\}_\theta$ are combined in algebraic quantities, in order to build different loss terms. These loss terms will be then summed to build the total loss which will guide the optimization NN parameters. With this benchmark study we want to investigate how different auxiliary loss terms affect the training process, comparing the performances on the common test dataset. Here we describe the different loss terms in detail.

\paragraph{Ground truth term.}
Following the MSE definition of Eq.~\ref{eq:data_loss}, we impose the ``ground truth'' loss term $\mathcal{L}_\mathrm{t}$ as 
\begin{equation}
    \mathcal{L}_\mathrm{t}(\theta)=\mathrm{MSE}\left(\mathcal{M}_\theta,\hat{\mathcal{M}}\right)+\mathrm{MSE}\left(M_{\mathrm{tot},\,\theta},\,\hat{M}_{\mathrm{tot}}\right)+\mathrm{MSE}\left(\eta_\theta,\hat{\eta}\right)\,.
    \label{eq:L_target}
\end{equation}
As we pointed out in Sec. \ref{sec:Auxiliary Physics-Informed Loss}, this is usually the basic loss term for most FCNNs. We choose this as the ``reference'' loss term for our NN, to which we will add and benchmark different APRIL terms.

\paragraph{Physical and algebraic terms.} 
The three quantities $\left\{\mathcal{M},M_\mathrm{tot},\eta\right\}$ are related thanks to Eq.~\ref{eq:mass_bound}. We can exploit this relation to calculate the following algebraic quantities between the NN outputs:
\begin{equation}
    \mathcal{M}_\mathrm{alg}=M_{\mathrm{tot},\,\theta}\,\eta_\theta^{3/5}\,,\qquad M_\mathrm{tot,\,alg}=\frac{\mathcal{M}_\theta}{\eta_\theta^{3/5}}\,,\qquad\eta_\mathrm{alg}=\left(\frac{\mathcal{M}_\theta}{M_{\mathrm{tot},\,\theta}}\right)^{5/3}\,.
\end{equation}
With these we define the ``physical'' $\mathcal{L}_\mathrm{p}$ and ``algebraic'' $\mathcal{L}_\mathrm{a}$ loss terms 
\begin{align}
\mathcal{L}_\mathrm{p}=&\mathrm{MSE}\left(\mathcal{M}_\mathrm{alg},\mathcal{M}_\theta\right)+\mathrm{MSE}\left(M_{\mathrm{tot,\,alg}},\,M_{\mathrm{tot},\,\theta}\right)+\mathrm{MSE}\left(\eta_\mathrm{alg},\eta_\theta\right)\,,\\    \mathcal{L}_\mathrm{a}=&\mathrm{MSE}\left(\mathcal{M}_\mathrm{alg},\hat{\mathcal{M}}\right)+\mathrm{MSE}\left(M_{\mathrm{tot,\,alg}},\,\hat{M}_{\mathrm{tot}}\right)+\mathrm{MSE}\left(\eta_\mathrm{alg},\hat{\eta}\right)\,.
\end{align}
In $\mathcal{L}_\mathrm{p}$ we are comparing the algebraic relations with the NN outputs, while in $\mathcal{L}_\mathrm{a}$ we are doing the same with the target features.

\paragraph{Frequency derivative term.}
The second relation we want to exploit is the $\tfrac{df}{dt}$ dependence. The three mass quantities $\left\{\mathcal{M},M_\mathrm{tot},\eta\right\}$ enters in the 1.5PN $\tfrac{df}{dt}$, as shown in Eq. \ref{eq:dfdt_pn_separation}. Substituting NN outputs $\left\{\mathcal{M},M_\mathrm{tot},\eta\right\}_\theta$ in these  functions, we define the quantities $\left.\frac{df}{dt}\right|_{\mathrm{Newt,\,alg}}$ and $\left.\frac{df}{dt}\right|_{\mathrm{corr,\,alg}}$. Thanks to these algebraic quantities, we introduce the following term:
\begin{equation}
    \mathcal{L}_\textrm{df}=\beta\mathcal{L}_\textrm{Newt}+\gamma\mathcal{L}_\textrm{corr}=\beta\,\,\textrm{MSE}\left(\left.\frac{df}{dt}\right|_{\mathrm{Newt,\,alg}},\left.\frac{df}{dt}\right|^\wedge_\mathrm{Newt}\right)+\gamma\,\,\textrm{MSE}\left(\left.\frac{df}{dt}\right|_{\mathrm{corr,\,alg}},\left.\frac{df}{dt}\right|^\wedge_\mathrm{corr}\right)\,,
    \label{eq:L_df}
\end{equation}
where $\beta$ and $\gamma$ are two fixed user-provided hyperparameters. Here, $\left.\frac{df}{dt}\right|^\wedge_\mathrm{Newt}$ and $\left.\frac{df}{dt}\right|^\wedge_\mathrm{corr}$ are the stored target, as described in Sec.~\ref{sec:generation_procedure}. Even for the algebraic quantities, the $f_k$ quantity in Eq.~\ref{eq:dfdt_newt_separation} is substituted directly with its target value $\hat{f}_k$. This is done because no analytical formula for $f(t;m_1,m_2)$ is available in literature for the 1.5PN formalism. Despite that, we choose to do this in order to still exploit the $\frac{df}{dt}$ dependence on mass parameters. 

We emphasize that the optimization is highly sensitive to the relative scale of the loss components: without reweighting, the term with the largest magnitude would dominate the gradients during backpropagation, while smaller-magnitude terms would be effectively ignored. For quantities in physical units, we find $\mathcal{L}_{\mathrm{Newt}}\sim \mathcal{O}(10^{3})$ and $\mathcal{L}_{\mathrm{corr}}\sim \mathcal{O}(10^{-4})$. Setting $\beta=10^{-6}$ and $\gamma=10^{4}$ brings the contributions to a comparable $\mathcal{O}(1)$ scale and ensures that each parameter has a similar influence on the learning of $df/dt$.

\paragraph{Total loss.} Summing together the loss terms, we define the NN total loss as
\begin{equation}
    \mathcal{L}_\mathrm{total}(\theta)=\alpha_\mathrm{t}\mathcal{L}_\mathrm{t}(\theta)+\alpha_\mathrm{APRIL}\mathcal{L}_\mathrm{APRIL}(\theta)=\alpha_\mathrm{t}\mathcal{L}_\mathrm{t}(\theta)+\alpha_\mathrm{APRIL}\left[\mathcal{L}_\mathrm{df}(\theta)+\mathcal{L}_\mathrm{p}(\theta)+\mathcal{L}_\mathrm{a}(\theta)\right]\,,
    \label{eq:L_total_bench}
\end{equation}
with $\left\{\alpha_\mathrm{t},\alpha_\mathrm{APRIL}\right\}$  denoting boolean hyper-parameters.

As humans we can interpret the output values in terms of their physical meaning, but for the machine $\mathcal{L}_\mathrm{t}(\theta)$ is only a numerical comparison between values. Given what has been said in \ref{sec:Auxiliary Physics-Informed Loss}, one can safely add APRIL terms combining a function of outputs with either outputs ($\mathcal{L}_\mathrm{p}$), targets ($\mathcal{L}_\mathrm{a}$) or a function of targets ($\mathcal{L}_\mathrm{df}$). This is fundamental to physics inform the network on how the outputs must be bound together: an information that is not implemented in $\mathcal{L}_\mathrm{t}$. Adding $\mathcal{L}_\mathrm{APRIL}$ we are really implementing the physical laws inside the training algorithm. In this context, it must be explicitly stated that we are not learning the ${df}/{dt}$ dynamics with a residual loss term as in usual hard-constrained PINNs \cite{RAISSI2019686}, but rather enforcing a pointwise mapping where the physical losses are consistency checks among different outputs.

\subsubsection{Loss terms benchmark}
\label{sec:Test_description}
In order to study the importance of physically-informed loss terms, we will compare the performance on $\left\{\mathcal{M},M_\mathrm{tot},\eta\right\}$ PE considering different combinations of loss terms. To do so, we perform 3 runs modifying the values of $\left\{\alpha_\mathrm{t},\alpha_\mathrm{APRIL}\right\} \in \left\{0,1\right\}$, i.e. either switch on or off contributions to the loss. The benchmark will be performed with different models, but with the same test dataset, as described in Sec.~\ref{sec:test_dataset}. In particular, we defined the Relative L1 (RL1) metric:
\begin{equation}
    \mathrm{RL1}\left(x_{\theta,\,i},\hat{x}_i\right)=\frac{1}{T}\sum_{i=1}^T\frac{\left|x_{\theta,\,i}-\hat{x}_i\right|}{\hat{x}_i}\,,
\end{equation}
where $\left\{x_{\theta,\,i}\right\}_{i=1}^T$ are the NN outputs, $\left\{\hat{x}_i\right\}_{i=1}^T$ are the target values, $x_i\in\mathbb{R}$ and $T=10^3$ is the test dataset dimension. While the training and validation datasets are passed as PyTorch datasets, the common test dataset is passed to the model as point-like measurements. This is done in order to simulate the real GW-event case, when events are analyzed one by one.

Addopting the RL1 metric results in the mean of the relative error arising in the test evaluation. It is more informative than a simple absolute error due to the difference in magnitude between the parameters that we are considering. The better the model will reproduce the target, the lower RL1 value will be.

For each of the 3 runs we apply RL1 to the three NN outputs $\left\{\mathcal{M},M_\mathrm{tot},\eta\right\}$ individually. The sum of the three parameters RL1 is also computed per run. We also repeat the 3 runs for 3 different training-validation datasets dimensions $D\in\left\{1,2,5\right\}\times10^3$ and for 3 batch sizes $B\in\left\{16,32,64\right\}$. This is done in order to study parameters which can affect the accuracy other than the losses themselves. Furthermore, we run the training 5 times changing the seed set before the training loop, to provide different learning paths for the same dataset. This results in 3 loss factors $\times$ 3 $D$ values  $\times$ 3 $B$ values $\times$ 5 seeds = 135 runs in total.

\section{Results and discussion}
\label{sec:discussion}
\begin{figure}[t!]
    \centering
    \includegraphics[width=1.\linewidth]{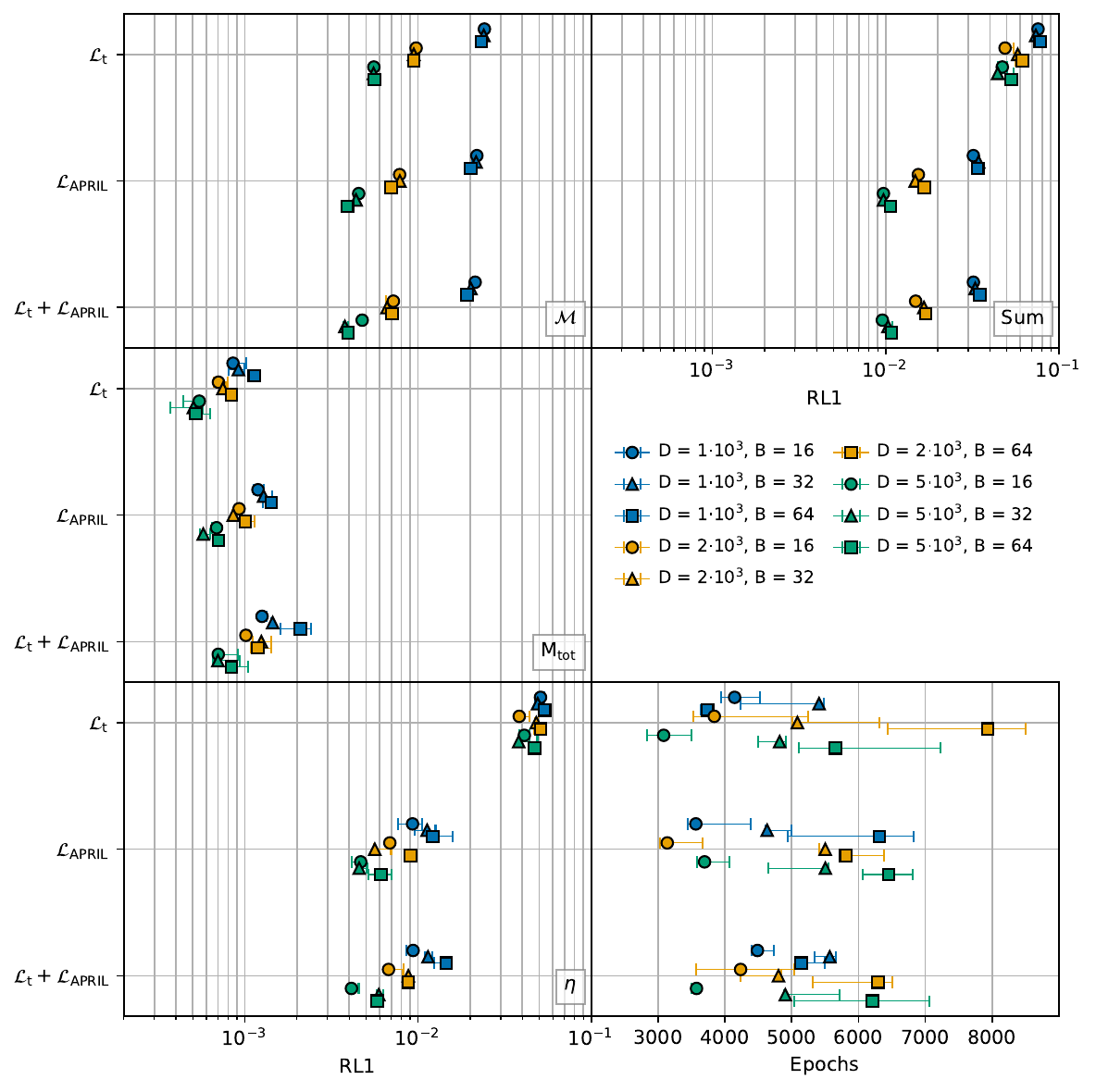}
    \caption{\emph{RL1 results (median and 68\% CI) for all the runs. Left panels show the study of the mass parameters individually, upper right panel shows the same study for the sum of the three, while bottom right panel shows the needed epochs to converge. The shape and the color of the markers determines the training dataset and the batch sizes. When APRIL losses are absent the RL1 result for the common sum is worse by a factor 10. Furthermore, the higher $D$ and the lower $B$, the better the results. Most of the APRIL impact is related to the $\eta$ term; see the text for discussion.}}
    \label{fig:all_runs_2}
\end{figure}
\begin{table}[t!]
    \small
    \centering
    \begin{tabular}{ccr|rrrr|r|r}
        \multicolumn{3}{c|}{\emph{Run}}&\multicolumn{4}{c|}{\emph{RL1 ($\times10^{-3}$)}}&\multicolumn{1}{c|}{\emph{R.}}&\multicolumn{1}{c}{\emph{Epochs}}\\
        \cmidrule{1-7}
        $D$&$B$&\multicolumn{1}{c|}{\emph{Loss terms}}&\multicolumn{1}{c}{$\mathcal{M}$}&\multicolumn{1}{c}{$M_\mathrm{tot}$}&\multicolumn{1}{c}{$\eta$}&\multicolumn{1}{c|}{\emph{Sum}}&&\\
        \toprule
        \multirow{9}{*}{$1\cdot10^3$}&\multirow{3}{*}{16}&$\mathcal{L}_\mathrm{t}$&$24.14^{+0.00}_{-0.00}$&$0.85^{+0.17}_{-0.00}$&$50.90^{+2.40}_{-1.19}$&$75.89^{+2.20}_{-1.31}$&26&$4146^{+377}_{-198}$\\
        &&$\mathcal{L}_\mathrm{APRIL}$&$21.84^{+0.29}_{-0.84}$&$1.19^{+0.10}_{-0.07}$&$9.29^{+1.29}_{-1.61}$&$32.12^{+1.45}_{-1.08}$&13&$3566^{+823}_{-120}$\\
        &&$\mathcal{L}_\mathrm{t}+\mathcal{L}_\mathrm{APRIL}$&$21.43^{+0.10}_{-0.69}$&$1.12^{+0.13}_{-0.04}$&$10.31^{+0.79}_{-0.50}$&$32.86^{+0.23}_{-0.19}$&14&$4488^{+240}_{-91}$\\
        \cmidrule{2-9}
        &\multirow{3}{*}{32}&$\mathcal{L}_\mathrm{t}$&$23.95^{+0.03}_{-0.08}$&$0.92^{+0.08}_{-0.11}$&$49.00^{+0.45}_{-1.46}$&$73.71^{+0.43}_{-1.42}$&25&$5410^{+78}_{-1171}$\\
        &&$\mathcal{L}_\mathrm{APRIL}$&$21.63^{+0.12}_{-1.03}$&$1.28^{+0.15}_{-0.06}$&$11.28^{+1.34}_{-1.72}$&$34.42^{+0.14}_{-1.63}$&16&$4633^{+365}_{-17}$\\
        &&$\mathcal{L}_\mathrm{t}+\mathcal{L}_\mathrm{APRIL}$&$21.16^{+0.07}_{-0.86}$&$1.31^{+0.09}_{-0.10}$&$11.96^{+1.30}_{-0.52}$&$34.54^{+0.35}_{-0.53}$&17&$5570^{+89}_{-231}$\\
        \cmidrule{2-9}
        &\multirow{3}{*}{64}&$\mathcal{L}_\mathrm{t}$&$23.16^{+0.19}_{-0.40}$&$1.13^{+0.00}_{-0.07}$&$53.71^{+0.10}_{-0.97}$&$78.14^{+0.00}_{-1.21}$&27&$3740^{+17}_{-107}$\\
        &&$\mathcal{L}_\mathrm{APRIL}$&$20.16^{+0.23}_{-0.72}$&$1.42^{+0.08}_{-0.15}$&$12.15^{+3.77}_{-0.51}$&$34.05^{+2.67}_{-0.12}$&15&$6312^{+513}_{-1374}$\\
        &&$\mathcal{L}_\mathrm{t}+\mathcal{L}_\mathrm{APRIL}$&$20.45^{+0.87}_{-0.18}$&$1.63^{+0.12}_{-0.31}$&$13.37^{+0.61}_{-1.97}$&$35.62^{+0.00}_{-0.95}$&18&$5142^{+355}_{-80}$\\
        \midrule
        \multirow{9}{*}{$2\cdot10^3$}&\multirow{3}{*}{16}&$\mathcal{L}_\mathrm{t}$&$9.74^{+0.12}_{-0.05}$&$0.70^{+0.09}_{-0.04}$&$38.37^{+5.71}_{-1.40}$&$48.95^{+5.77}_{-1.50}$&21&$3842^{+1398}_{-312}$\\
        &&$\mathcal{L}_\mathrm{APRIL}$&$7.82^{+0.00}_{-0.23}$&$0.93^{+0.00}_{-0.02}$&$6.87^{+0.12}_{-0.21}$&$15.46^{+0.10}_{-0.26}$&8&$3138^{+535}_{-110}$\\
        &&$\mathcal{L}_\mathrm{t}+\mathcal{L}_\mathrm{APRIL}$&$7.57^{+0.19}_{-0.05}$&$0.87^{+0.00}_{-0.04}$&$8.09^{+0.09}_{-0.85}$&$16.41^{+0.16}_{-0.56}$&9&$4236^{+801}_{-667}$\\
        \cmidrule{2-9}
        &\multirow{3}{*}{$32$}&$\mathcal{L}_\mathrm{t}$&$9.44^{+0.03}_{-0.04}$&$0.75^{+0.03}_{-0.00}$&$48.14^{+0.93}_{-0.86}$&$57.93^{+1.41}_{-0.65}$&23&$5086^{+1228}_{-73}$\\
        &&$\mathcal{L}_\mathrm{APRIL}$&$7.85^{+0.10}_{-0.00}$&$0.86^{+0.03}_{-0.02}$&$5.62^{+1.36}_{-0.07}$&$14.83^{+0.36}_{-0.37}$&7&$5503^{+11}_{-91}$\\
        &&$\mathcal{L}_\mathrm{t}+\mathcal{L}_\mathrm{APRIL}$&$7.59^{+0.47}_{-0.10}$&$0.82^{+0.15}_{-0.00}$&$8.13^{+0.14}_{-1.19}$&$16.53^{+0.05}_{-0.40}$&10&$4803^{+40}_{-574}$\\
        \cmidrule{2-9}
        &\multirow{3}{*}{$64$}&$\mathcal{L}_\mathrm{t}$&$9.44^{+0.18}_{-0.39}$&$0.83^{+0.02}_{-0.02}$&$50.89^{+0.29}_{-4.18}$&$61.51^{+0.00}_{-4.94}$&24&$7933^{+569}_{-1490}$\\
        &&$\mathcal{L}_\mathrm{APRIL}$&$6.98^{+0.14}_{-0.08}$&$1.01^{+0.13}_{-0.02}$&$9.03^{+0.14}_{-0.22}$&$16.70^{+0.50}_{-0.10}$&12&$5811^{+577}_{-69}$\\
        &&$\mathcal{L}_\mathrm{t}+\mathcal{L}_\mathrm{APRIL}$&$7.44^{+0.14}_{-0.44}$&$1.21^{+0.00}_{-0.03}$&$8.40^{+0.23}_{-0.60}$&$16.58^{+1.02}_{-0.14}$&11&$6292^{+210}_{-981}$\\
        \midrule
        \multirow{9}{*}{$5\cdot10^3$}&\multirow{3}{*}{16}&$\mathcal{L}_\mathrm{t}$&$5.55^{+0.15}_{-0.03}$&$0.55^{+0.02}_{-0.10}$&$41.06^{+2.97}_{-2.77}$&$47.17^{+2.90}_{-2.75}$&20&$3084^{+416}_{-244}$\\
        &&$\mathcal{L}_\mathrm{APRIL}$&$4.54^{+0.16}_{-0.10}$&$0.69^{+0.00}_{-0.04}$&$4.66^{+0.05}_{-0.51}$&$9.72^{+0.15}_{-0.16}$&1&$3696^{+376}_{-109}$\\
        &&$\mathcal{L}_\mathrm{t}+\mathcal{L}_\mathrm{APRIL}$&$4.29^{+0.30}_{-0.20}$&$0.61^{+0.02}_{-0.08}$&$5.90^{+0.22}_{-0.57}$&$10.45^{+0.17}_{-0.12}$&4&$3576^{+26}_{-71}$\\
        \cmidrule{2-9}
        &\multirow{3}{*}{$32$}&$\mathcal{L}_\mathrm{t}$&$5.53^{+0.17}_{-0.00}$&$0.50^{+0.07}_{-0.13}$&$38.09^{+10.70}_{-0.11}$&$44.37^{+10.39}_{-0.41}$&19&$4820^{+89}_{-318}$\\
        &&$\mathcal{L}_\mathrm{APRIL}$&$4.38^{+0.04}_{-0.11}$&$0.58^{+0.05}_{-0.02}$&$4.57^{+0.50}_{-0.03}$&$9.72^{+0.15}_{-0.05}$&2&$5505^{+51}_{-850}$\\
        &&$\mathcal{L}_\mathrm{t}+\mathcal{L}_\mathrm{APRIL}$&$4.38^{+0.10}_{-0.18}$&$0.66^{+0.08}_{-0.00}$&$4.93^{+1.01}_{-0.07}$&$10.25^{+0.56}_{-0.28}$&3&$4905^{+809}_{-19}$\\
        \cmidrule{2-9}
        &\multirow{3}{*}{$64$}&$\mathcal{L}_\mathrm{t}$&$5.59^{+0.02}_{-0.03}$&$0.52^{+0.11}_{-0.03}$&$47.00^{+1.35}_{-1.38}$&$53.29^{+1.58}_{-1.65}$&22&$5657^{+1576}_{-552}$\\
        &&$\mathcal{L}_\mathrm{APRIL}$&$3.91^{+0.31}_{-0.12}$&$0.71^{+0.03}_{-0.00}$&$6.10^{+0.96}_{-0.92}$&$10.69^{+0.85}_{-0.57}$&5&$6454^{+364}_{-384}$\\
        &&$\mathcal{L}_\mathrm{t}+\mathcal{L}_\mathrm{APRIL}$&$4.17^{+0.24}_{-0.16}$&$0.66^{+0.05}_{-0.00}$&$5.94^{+0.52}_{-0.48}$&$10.86^{+0.35}_{-0.31}$&6&$6210^{+851}_{-1175}$\\
        \bottomrule
    \end{tabular}
    \caption{\emph{RL1 performance on the test dataset for the different models. The values corresponds to the median and the $68\%$ CI and are plotted in Fig.~\ref{fig:all_runs_2}.}}
    \label{tab:test_runs_2}
\end{table}
The RL1 results for the common test dataset are summarized in Fig.~\ref{fig:all_runs_2}, with corresponding values listed in Table~\ref{tab:test_runs_2}. Each panel represents the results for the mass parameters depending on the run, and on the training data size $D$ (marker color) and batch size $B$ (marker shape). Every point (errorbar) corresponds to the median (68\% Confidence Interval, CI) of the 5 training repetitions. The different runs are ranked in Table~\ref{tab:test_runs_2} based on the "$Sum$" RL1 median, which is informative on the overall performance of the model  since RL1 reflects a relative error on parameters.
\par Before going into the details of the results, we report that the computational time for the runs depends mostly on $D$ and it results to be $\sim0.33$ hr for $D=10^3$, $\sim1$ hr for $D=2\cdot10^3$ and $\sim2$ hr for $D=5\cdot10^3$ on NVIDIA GeForce RTX 4060 \cite{NvidiaRTX4060} and 4070 \cite{NvidiaRTX4070} GPUs (without parallelization inside the training loop).
\par  Extended results benchmarking all different $\mathcal{L}_\mathrm{APRIL}$ terms can be seen in Fig.~\ref{tab:test_runs}, and Fig.~\ref{fig:all_runs} in Appendix \ref{app:extended_results}.

\subsection{Impact of $\mathcal{L}_\mathrm{APRIL}$ on parameter accuracy}

Looking at the "Sum" panel in Fig.~\ref{fig:all_runs_2} and focusing on one value of $D$ and $B$, one can clearly see that when physical redundant losses $\mathcal{L}_\mathrm{APRIL}$ are present, the RL1 results are systematically lower. Notice that this is not the same for every single mass component: for $\mathcal{M}$ (upper left panel) the results are similar, but less relevant, while for $M_\mathrm{tot}$ (center left panel), beside the overall agreement with other results, the $\mathcal{L}_\mathrm{t}$ run is the best one for all the $\left\{D,B\right\}$ pairs. Anyway, the better performance when considering the sum of all three parameter relative errors is present. This happens because $\mathcal{L}_\mathrm{APRIL}$ plays a key role in the learning of $\eta$ (lower-left panel). By definition (Eq.~\ref{eq:mass_related}), and due to its magnitude and the shape of the test dataset (Fig.~\ref{fig:trainval_test}), $\eta$ is inherently difficult to learn. In this context, the redundant loss terms provide a benefit: the learning of $\mathcal{M}$ and $M_\mathrm{tot}$ supports the learning of $\eta$ through $\mathcal{L}_\mathrm{df}$, $\mathcal{L}_\mathrm{p}$, and $\mathcal{L}_\mathrm{a}$, leading to improved accuracy for $\eta$ compared to training without them. The RL1 sum was selected as the ranking metric precisely because it highlights the overall learning process. Furthermore, notice that even the run with only $\mathcal{L}_\mathrm{APRIL}$ has a lower RL1 "Sum" value than runs with $\mathcal{L}_\mathrm{t}$ alone, confirming the validity of this approach.
\begin{figure}[t!]
    \centering
    \includegraphics[width=1.\linewidth]{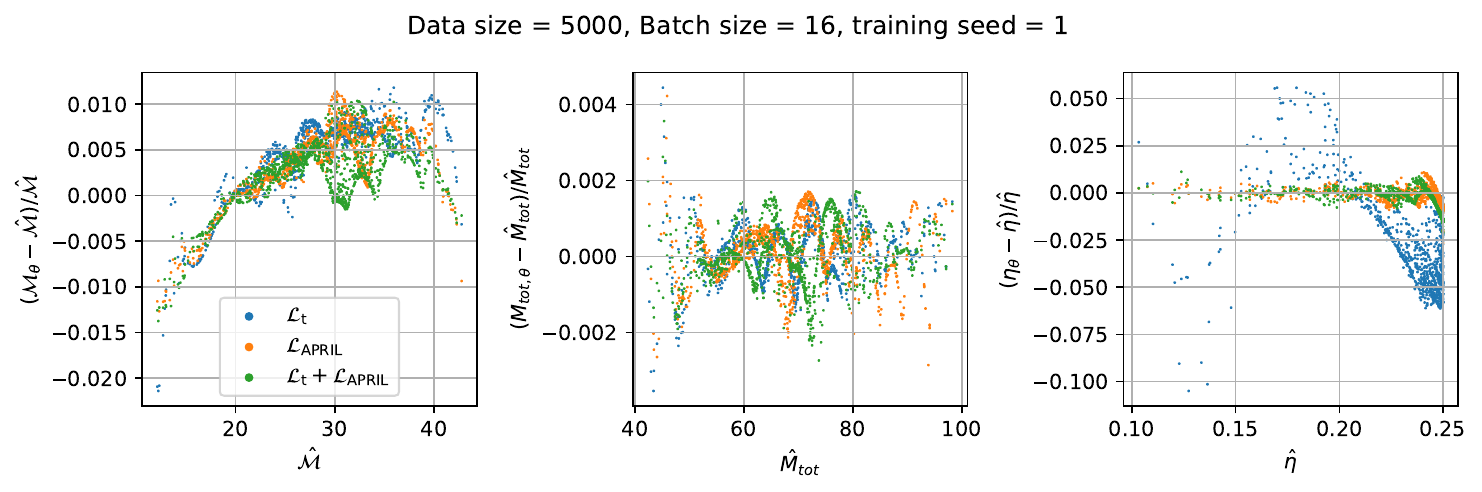}
    \caption{\emph{Test output relative errors on mass components comparing the runs for $\left\{D,B,\text{seed}\right\}=\left\{5\times10^3,16,1\right\}$. The results for the two runs with active APRIL are similar to $\mathcal{L}_\mathrm{t}$ for $\mathcal{M}$ and $M_\mathrm{tot}$, but are more accurate for $\eta$, spanning only 3\% in relative error instead of 10\%.}}
    \label{fig:pinn_runs_5000}
\end{figure}

The impact of $\mathcal{L}_\mathrm{APRIL}$ can be also seen in Fig.~\ref{fig:pinn_runs_5000}. Here, we plot the NN prediction output, i.e. relative residuals for the common test dataset parameter values, testing the models from different training runs. In particular we are looking at the case $\left\{D,B,\text{seed}\right\}=\left\{5\times10^3,16,1\right\}$, which gives the best runs of the whole study, but the same reasoning can be extended to every case. As we already saw in Fig.~\ref{fig:all_runs_2}, there is no relevant difference for $\mathcal{M}$ and $M_\mathrm{tot}$, while the improvement is evident for $\eta$. For the latter, $\mathcal{L}_\mathrm{t}$ spans a relative error of $10\%$, while the other two reach a maximum disagreement of $3\%$. The presence of the redundant losses helps to train $\eta$ thanks to the implemented bound with $\mathcal{M}$ and $M_\mathrm{tot}$. The key result of this study is that including the physical redundant terms $\mathcal{L}_\mathrm{df}$, $\mathcal{L}_\mathrm{p}$ and $\mathcal{L}_\mathrm{a}$ balances the learning of the different parameters, resulting in better overall agreement with the target values of all mass components simultaneously. This balancing can be directly attributed to the enforcement of physical relations within the loss function itself, which guides the training algorithm toward physically meaningful minima, as demonstrated in Sec.~\ref{sec:global_minimum_effects}.

\par In the context of Fig.~\ref{fig:pinn_runs_5000}, we  also highlight the strong points of our approach. Even if the training was done on a dataset generated from $M_\mathrm{tot}$ and $\eta$ uniform sampling, the application to a dataset generated from a different distribution retains good performance inside the same mass domains. Training with uniform sampling avoids the formation of value clusters near critical points, like for example we can see in \cref{fig:trainval_test} for the GWTC-4 distribution, and thus ensures more uniform training across the entire parameter space.

\subsection{Importance of $\mathcal{L}_\mathrm{APRIL}$ term on the training process}
\begin{figure}[t!]
    \centering
    \includegraphics[width=1.\linewidth]{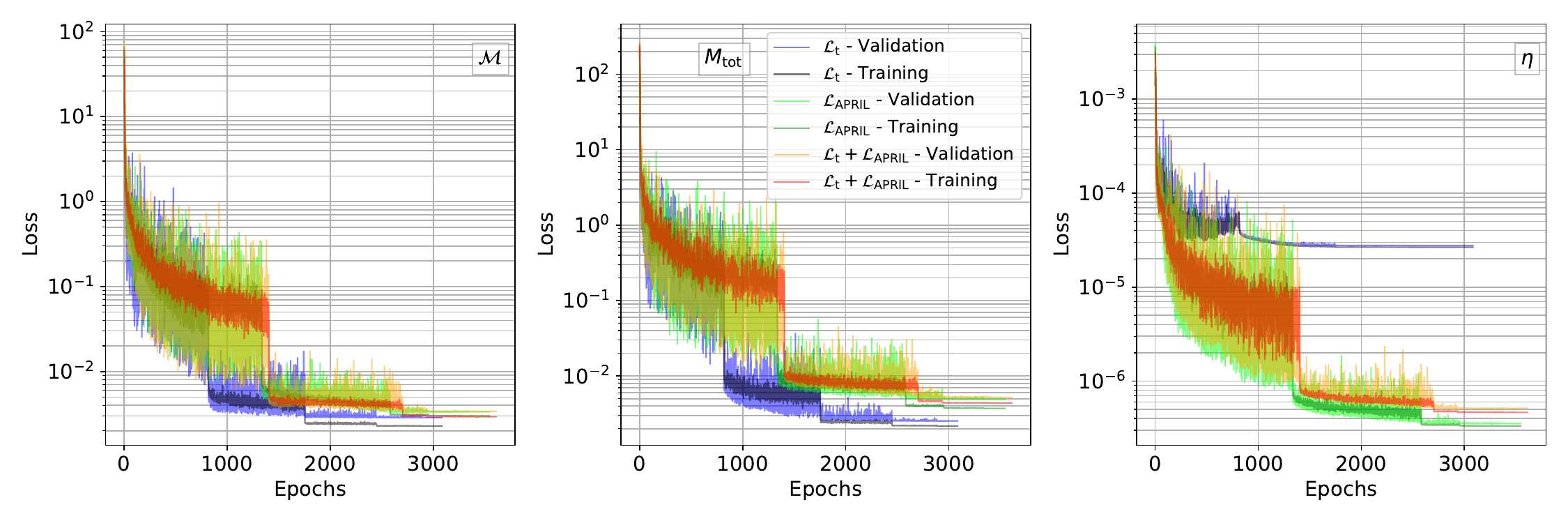}
    \caption{\emph{Ground truth loss components for the different runs. The combination of hyperparameters for these runs are $\left\{D,B,\text{seed}\right\}=\left\{5\times10^3,16,1\right\}$. Even if not included in $\mathcal{L}_\mathrm{APRIL}$ run, we plotted the single ground truth terms for every run as a metric to compare outputs of the NN with their target during the run. One can clearly see how there is no relevant difference for $\mathcal{M}$ and $M_\mathrm{tot}$, while again for $\eta$ the improvement is evident. Despite being more lengthy, the training for the runs with $\mathcal{L}_\mathrm{APRIL}$ is following a different loss landscape, where the absolute physical minimum is the same, but the gradient is augmented by the presence of new APRIL terms. This is translated to a better convergence towards a minimum with more physical sense than in the $\mathcal{L}_\mathrm{t}$ run.}}
    \label{fig:losses}
\end{figure}

In Fig.~\ref{fig:losses} one can see the values for the ground truth loss components (Eq.~\ref{eq:L_target}) during the runs, which were also computed for $\mathcal{L}_\mathrm{APRIL}$, but not included. These terms can serve as a metric to compare the output values with their target during the training loops. For coherence, the $\left\{D,B,\text{seed}\right\}$ combination is the same as the last paragraph.

Once more it is shown how the $\mathcal{L}_\mathrm{APRIL}$ losses are guiding the algorithm towards physical minima. While for $\mathcal{M}$ and $M_\mathrm{tot}$ the difference is not very relevant, one can see the two order of magnitude improvement for $\eta$. The initial convergence is a bit harder, but the $\eta$ agreement suggests that the algorithm is driven towards a physical minimum. Despite being more lengthy, the training for the two runs with $\mathcal{L}_\mathrm{APRIL}$ is following a different loss landscape, where the absolute physical minimum is the same, but the gradient is augmented by the presence of APRIL new terms. This is perfectly in agreement with the demonstration in Sec.~\ref{sec:global_minimum_effects}, and translates to a better convergence with more physical sense than for the $\mathcal{L}_\mathrm{t}$ run. 

\subsection{Dependence of parameter's accuracy on the training data size}
\begin{figure}[t!]
    \centering
    \includegraphics[width=1\linewidth]{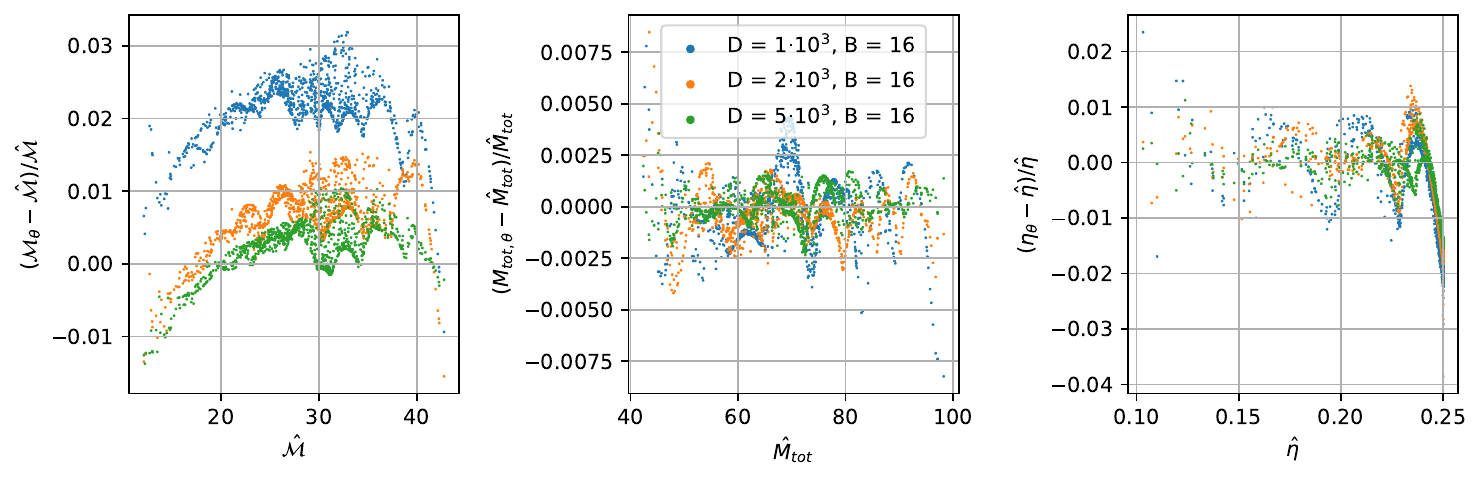}
    \caption{\emph{RL1 error comparing $\mathcal{L}_\mathrm{t}+\mathcal{L}_\mathrm{APRIL}$ for the same batch size and a different data size. Here we set always the seed value to $1$. We can see clearly a shift for $\mathcal{M}$, which is anyway compensated with increasing $D$.}}
\label{fig:pinn_runs_best_samebatch}
\end{figure}
\par Looking again at Fig.~\ref{fig:all_runs_2}, one can see a dependence on the data size $D$: there is a clear systematic pattern in favor of a bigger dataset. This is an expected behavior when dealing with a training algorithm: the higher the dataset dimension, the more steps the optimization process performs per epoch. The "Epochs" panel shows the same pattern: the median number of epochs decreases with increasing $D$.

This reasoning is related to the $\mathcal{M}$ results, which depends strongly on $D$. To explain this, we plot in Fig.~\ref{fig:pinn_runs_best_samebatch} the mass results for $\mathcal{L}_\mathrm{t}+\mathcal{L}_\mathrm{APRIL}$, varying only the data size $D$, with a fixed training seed. One can see the difference between the runs in the $\mathcal{M}$ plot. For a lower $D$ we have a lower accuracy due to a systematic shift in the prediction of the order of $2\%$. The fact that this is common to all runs for the same $D$ suggests that this shift is caused by $\mathcal{M}$ influencing the main Newtonian component of the signal, being the most sensitive mass term among the three, and prone to overfitting. $\mathcal{L}_\mathrm{APRIL}$ terms are not able to balance the whole magnitude of the shift, but the $\mathcal{M}$ panel of Fig.~\ref{fig:all_runs_2} shows some improvement in that respect. This improvement is also visible in results for the GWTC-4 distribution, which are sufficiently more complex than the training dataset.

\section{Conclusions}
\label{sec:conclusions}

We have studied here auxiliary physically-redundant information included in loss (APRIL) as additive components that exploit known physical relationships among NN outputs. We mathematically demonstrated the validity of this approach and performed benchmark runs for different loss-term combinations using a deliberately simple GW PE approach with a NN. Both the theoretical analysis and the benchmark results show that including the APRIL terms guide the training toward physically meaningful minima by increasing the curvature of the loss landscape around them, thereby improving accuracy and enforcing physical consistency. Even simple known physical relations are sufficient to increase by an order-of-magnitude the overall test accuracy. In particular, the benchmark study highlights that APRIL terms enhance the estimation of parameters that are otherwise too difficult to learn.

While this approach is not intended to compete with strong-form PDE solving PINNs, it represents a powerful way of combining physics-information inside the loss function and the simultaneous PE for many realization of the same physical system. Despite being applied only to a GW case, we believe that this approach can be suitable for many different fields of science and engineering, being only based on existing relations between output features.

A key element of this work is the introduction of a NN for GW PE, inspired by the PINNs framework. Using simulated GW frequency signals, we successfully estimated multiple mass-related quantities with a good degree of interpolation between different mass distributions. In the present context, the network serves as a proof of concept for benchmarking APRIL, but it could also form the basis for more complex physics-inspired architectures applied directly to GW data analysis. The light PINN-inspired approach presented here could be developed into a practical PE tool for unmodeled pipelines in both current ground-based interferometers and the upcoming Einstein Telescope, with the inclusion of physical constraints compensating for the minimal assumptions typically used in detection.
\par Future work will aim to integrate this APRIL approach more directly into detection pipelines, for example by training on time-frequency spectrograms rather than simulated frequency series, enabling the network to operate directly on detection outputs. For this purpose, the network will need to be fine-tuned and studied adding noise and systematically varying the signal-to-noise ratio. GW science turned to be particularly suitable for APRIL because of the physical redundancy between mass terms and the structure of the Post-Newtonian (PN) expansion, which isolates parameter contributions at different orders. Expanding the PN order would naturally increase the number of parameters to be estimated. Extending the mass range to ${\sim}3\,M_{\odot}$ would include the main BH population peak as well as NS-BH systems. These developments will be part of a future effort to evolve this proof of concept into the first PINN-inspired GW PE pipeline.

\section*{Acknowledgments}

This research was partially supported by the European Union-Next Generation EU, Mission 4 Component 1 CUP J53D23001550006 with the PRIN Project No.~202275HT58; by the Polish National Science Center OPUS grant no. 2021/43/B/ST9/01714; by ICSC – Centro Nazionale di Ricerca in High Performance Computing, Big Data and Quantum Computing, funded by European Union – NextGenerationEU; and by the National Aeronautics and Space Administration (NASA) under Award 80NSSC24K0767. We acknowledge using computing resources of the Ferrara section of National Institute of Nuclear Physics (INFN), Italy and of the Nicolaus Copernicus Astronomical Center, Poland. As test dataset, in this paper we make use of GWTC-4 population data as published in \cite{theligoscientificcollaboration2025gwtc40populationpropertiesmerging,GWTC4_population_2025}. This material is based upon work supported by NSF's LIGO Laboratory which is a major facility fully funded by the National Science Foundation.

\appendix

\section{Robustness against noisy data}
\label{app:noise}
\begin{figure}[t!]
    \centering
    \includegraphics[width=0.88\linewidth]{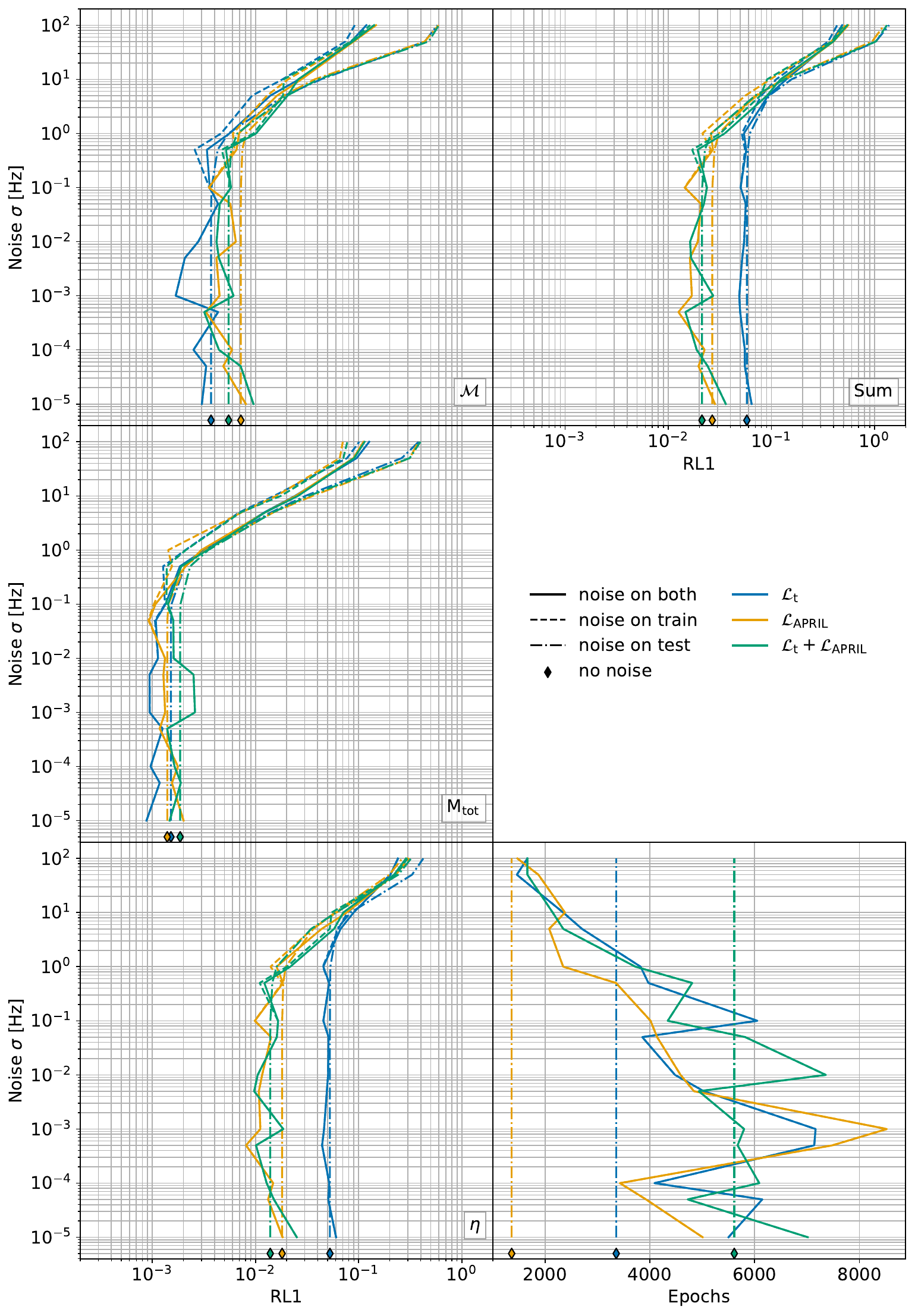}
    \caption{\emph{RL1 results for a single $\left\{D,B\right\}=\left\{10^3,64\right\}$ run with different levels of Gaussian noise $\mathcal{N}(\mu,\sigma)$, with $\mu=0$}\emph{, added to the frequency array $f_k$. The different panels present the study of the mass parameters individually, while the color of the lines determines the loss terms considered, and the different line styles indicates the set where the noise was added. At the bottom of each panel, the results on clean data (Sec.~\ref{sec:discussion}) are plotted as colored diamonds. In the "Sum" plot, one can clearly notice that the better accuracy when dealing with $\mathcal{L}_\mathrm{APRIL}$ and $\mathcal{L}_\mathrm{t}+\mathcal{L}_\mathrm{APRIL}$ proceeds from the clear case up to $\sigma=5$} Hz\emph{. Coherently with the main results, the better overall agreement comes from learning $\eta$, balancing the learning among other parameters. Furthermore, as one should expect, the results for the tests on noisy signals trained with the clear dataset stick roughly at the same RL1 value up to $\sigma=0.1$} Hz\emph{, since they come from the same training process.}}
\label{fig:pinn_runs_noise}
\end{figure}
In the main text we introduced APRIL as the approach of adding to the supervised loss function different terms which exploit the physical redundancy between the output parameters describing the data. The analysis was focused on simulated clean (``noiseless'') signals purely defined by the mass parameters. Here we extend the study to noisy signals.

The new signals can be expressed, for each time bin $t_k$, as the sum of the default simulated 1.5PN signal $f_k$ and a randomly-drawn value from a Gaussian noise distribution $n_k=\mathcal{N}_k(\mu,\sigma)$:
\begin{equation}
    f_{k,\textrm{noisy}} = f_k + n_k = f_k+\mathcal{N}_k(\mu,\sigma)\,,
\end{equation}
with the mean value $\mu=0$. With this choice, we can test the robustness of our APRIL approach. Indeed, while $f_k$ is still dependent on the $\left\{\mathcal{M},M_\mathrm{tot},\eta\right\}$ parameters as expressed in \cref{sec:physical_context}, $n_k$ is not.

We generate 15 train and 15 test noisy datasets sampling from two different distributions (uniform and GWTC-4 inferred respectively, as described in Sec.~\ref{sec:generation_procedure}), with the $\sigma$ value changing for different datasets. In particular, we consider $\sigma$ in the range of $\left[10^{-5},\,10^{2}\right]$ Hz. The data seed is set before the sampling loop, leading to the same noise realizations among different levels of $\sigma$. Different seeds are set for the train and the test datasets.

Using these datasets and the default clean datasets, with the runs hyperparameters $\left\{D,B\right\}=\left\{10^3,64\right\}$, we explore four cases:
\begin{itemize}
    \item \emph{No noise}, i.e. the default Sec.~\ref{sec:Test_description} runs trained and tested on the clean test datasets;
    \item \emph{Noise on test}, where we take the Sec.~\ref{sec:Test_description} runs trained on the clean training dataset, without repeating them, and test them on the new test datasets;
    \item \emph{Noise on train}, where we train new models on the 15 noisy train datasets and we test them on the clean test dataset;
    \item \emph{Noise on both}, where we test the previous point models with the new test datasets; in this case, the ratio between the train and the test $\sigma$ values equals 1.
\end{itemize}
For each of these cases, we train three different models with the three combinations of the loss components.

The RL1 results are shown in Fig.~\ref{fig:pinn_runs_noise}, where different RL1 values are plotted against the selected $\sigma$ value. As usual, both single parameters and the overall sum results are present. One can distinguish among loss terms runs based on the line color, while the line style highlights the study case among the listed ones. In particular, the "\emph{No noise}" case is plotted as a single point at the bottom of each panel.

The main result is that the $\mathcal{L}_\mathrm{APRIL}$ and the $\mathcal{L}_\mathrm{APRIL}+\mathcal{L}_\mathrm{t}$ runs perform better than the $\mathcal{L}_\mathrm{t}$ run up to a noise level as high as $\sigma \approx 10$ Hz, while $f_k$ spans from $\mathcal{O}(10)$ Hz to $\mathcal{O}(10^2)$ Hz. As usual, the better overall behavior of the two runs with $\mathcal{L}_\mathrm{APRIL}$ comes from the action on the $\eta$ parameter, which is usually a stiff parameter to learn. $\mathcal{L}_\mathrm{t}$ is still individually better for $\mathcal{M}$ and $M_\mathrm{tot}$, since it optimizes one parameter per time with dedicated loss terms. The latter are precisely introducing the redundancy needed in order to optimize all the parameters jointly. Even if the performance on $\mathcal{M}$ and $M_\mathrm{tot}$ is sometimes worse, the redundancy stemming from these parameters is helping $\eta$ to be learned better thanks to the other two parameters, resulting in a better overall agreement, as we can see in the "$Sum$" results.

The results coming from "\emph{Noise on test}" runs are of particular interest for two reasons. First, they provide a consistency test. Indeed, for increasing $\sigma$ up to $\sim0.5$ Hz, the RL1 results don't vary. This is an expected behavior, since the model is trained only once and tested on different datasets. Since it was trained on noiseless signals, the addition of different levels of noise in the test dataset is not expected to have a better performance than the test on the clear dataset. On the other hand, when the noise level is sufficiently low, it doesn't impact the results with respect to the "\emph{No noise}" case. Second, "\emph{Noise on test}" performs usually worse than "\emph{Noise on train}" and "\emph{Noise on both}". One can observe this as a common characteristic among all different runs, more enhanced when $\sigma>0.1$. This has a simple meaning which is consistent with the generalization strategy that is usually implemented: training with a noise component permits to learn slightly better than doing the same with a clean signal, since training on noisy data is giving the needed amount of randomness to generalize better.

\section{Broken Power Law + 2 Peaks black hole distribution from GWTC-4 event catalog}
\label{app:03_BPL2P}
Here we describe the details of the common test dataset mass distribution. In \cite{theligoscientificcollaboration2025gwtc40populationpropertiesmerging}, the LVK collaboration released their study on the BBH population based on the GWTC-4 event catalog \cite{theligoscientificcollaboration2025gwtc40updatinggravitationalwavetransient}. The most likely distribution for the main mass component $m_1$ was found to be a broken power-law with 2 peaks (BPL2P):
\begin{equation}
    \begin{split}
        \pi\left(m_1|\mathbf{\Lambda}_\mathrm{m_1}\right)\propto&\Bigl[\lambda_0\mathcal{P}_\mathrm{BP}\left(m_1|\alpha_1,\alpha_2,m_\mathrm{break},m_\mathrm{1,low},m_\textrm{high}\right)+\lambda_1\mathcal{N}_\mathrm{lt}\left(m_1|\mu_1,\sigma_1,\mathrm{low}=m_\mathrm{1,low}\right)\\
        &+\left(1-\lambda_0-\lambda_1\right)\mathcal{N}_\mathrm{lt}\left(m_1|\mu_1,\sigma_1,\mathrm{low}=m_\mathrm{1,low}\right)\Bigr]S\left(m_1|m_\mathrm{1,low},\delta_{m,1}\right)
    \end{split}\,,
    \label{eq:p(m1)}
\end{equation}
where $\mathbf{\Lambda}_{m_1}=\left\{\alpha_1,\alpha_2,m_\mathrm{break},\mu_1,\sigma_1,\mu_2,\sigma_2,m_\mathrm{1,low},\delta_{m,1},\lambda_0,\lambda_1,m_\mathrm{high}\right\}$ are parameters listed in Table~\ref{tab:params_m1}.  $\mathcal{P}_\mathrm{BP}\left(m_1|\alpha_1,\alpha_2,m_\mathrm{break},m_\mathrm{1,low},m_\textrm{high}\right)$ is a normalized broken power-law distribution with spectral indexes $\left\{-\alpha_1,-\alpha_2\right\}$, $\mathcal{N}_\mathrm{lt}\left(m_1|\mu_1,\sigma_1,\mathrm{low}=m_\mathrm{1,low}\right)$ is a left-truncated normal distribution with mean $\mu_m$ and width $\sigma_m$, $\lambda_0$ and $\lambda_1$ are mixing fractions determining the relative prevalence of mergers in $\mathcal{P}$, $\mathcal{N}_\mathrm{lt,1}$ and $\mathcal{N}_\mathrm{lt,2}$, and $S(m_1|m_\mathrm{min},\delta_m)$ is a smoothing function, which rises from 0 to 1 over the interval $(m_\mathrm{min},m_\mathrm{min}+\delta_m)$. 
\begin{table}[t!]
    \small
    \centering
    \begin{tabular}{r|cccccccccccc}
        \emph{Parameter} & $\alpha_1$ & $\alpha_2$ & $m_\textrm{break}$ & $\mu_1$ & $\sigma_1$ & $\mu_2$ & $\sigma_2$ & $m_\mathrm{1,low}$ & $\delta_\mathrm{m,1}$ & $\lambda_0$ & $\lambda_1$ & $m_\mathrm{high}$\\
        \midrule
        \emph{Value} & 1.72 & 4.51 & 35.62 & 9.76 & 0.68 & 32.76 & 3.92 & 5.06 & 4.32 & 0.36 & 0.59 & 300.00\\
        \bottomrule
        \multicolumn{1}{c}{}
    \end{tabular}
    \caption{\emph{BPL2P parameter values for $\pi(m_1|\mathbf{\Lambda}_{m_1})$ in Eq.~\ref{eq:p(m1)}. $\alpha_1$, $\alpha_2$, $\lambda_0$ and $\lambda_1$ are unitless  quantities, while all the other have units of $M_\odot$.}}
    \label{tab:params_m1}
\end{table}
\begin{table}[t!]
    \centering
    \begin{tabular}{r|ccc}
        \emph{Parameter} & $\beta_\mathrm{q}$ & $m_\mathrm{2,low}$ & $\delta_\mathrm{m,2}$\\
        \midrule
        \emph{Value} & 1.17 & 3.55 & 4.91\\
        \bottomrule
    \end{tabular}
    \caption{\emph{BPL2P parameter values for $\pi(q|\mathbf{\Lambda}_\mathrm{q})$ in Eq.~\ref{eq:p(q)}. $\beta_\mathrm{q}$ is an unitless quantity, while $m_\mathrm{2,low}$ and $\delta_\mathrm{m,2}$ have units of M$_\odot$.}}
    \label{tab:params_q}
\end{table}
In particular, we used
\begin{align}
\mathcal{P}\left(m_1|\alpha_1,\alpha_2,m_\mathrm{break},m_\mathrm{high}\right)=&\frac{1}{N}
    \begin{cases}
        \left(\frac{m_1}{m_\mathrm{break}}\right)^{-\alpha_1}&m_\mathrm{1,low}\leq m_1<m_\mathrm{break}\\
        \left(\frac{m_1}{m_\mathrm{break}}\right)^{-\alpha_2}&m_\mathrm{break}\leq m_1<m_\mathrm{high}
    \end{cases}\,,\\
    \mathcal{N}_\mathrm{lt}\left(m_1|\mu,\sigma,\mathrm{low}=m_\mathrm{1,low}\right)=&
    \begin{cases}
        \frac{1}{\sigma}\frac{\phi\left(\frac{m_1-\mu}{\sigma}\right)}{1-\Phi\left(\frac{m_1-\mu}{\sigma}\right)}&m_1\geq m_\mathrm{1,low}\\
        0&m_1<m_\mathrm{1,low}
    \end{cases}\,,\\
    S\left(m_1|m_\mathrm{min},\delta_m\right)=&
    \begin{cases}
        0&m_1<m_\mathrm{min}\\
        \left[f\left(m_1-m_\mathrm{min},\delta_m\right)+1\right]^{-1}&m_\mathrm{min}\leq m_1\leq m_\mathrm{min}+\delta_m\\
        1&m_1\geq m_\mathrm{min}+\delta_m
    \end{cases}\,,
\end{align}
where
\begin{align}
    \qquad\phi\left(z\right)&=\frac{1}{\sqrt{2\pi}}\exp\left(\frac{-z^2}{2}\right)\,,\\\qquad\Phi(z)&=\int^z_{-\infty}\phi(t)dt\,,\\
    f\left(m',\delta_m\right)&=\exp\left(\frac{\delta_m}{m'}+\frac{\delta_m}{m'-\delta_m}\right)\,.
\end{align}
\par Once sampled $m_1$, the conditional probability for $q$ is given by
\begin{equation}
    \pi\left(q|\mathbf{\Lambda}_\mathrm{q}\right)\propto q^{\beta_\mathrm{q}}S(m_1q|m_\mathrm{2,low},\delta_{m,2})\,,
    \label{eq:p(q)}
\end{equation}
where $\mathbf{\Lambda}_\mathrm{q}=\left\{m_1,\beta_\mathrm{q},m_\mathrm{2,low},\delta_{m,2}\right\}$, listed in Table~\ref{tab:params_q}. The values for the parameters were taken from \cite{GWTC4_population_2025}. In particular we downloaded the data contained in the \cite{analyses_BBH.tar} folder and we took the median value for the given samples of each parameter, using the model highlighted as most likely \cite{BBHMassSpinRedshift}.

\begin{figure}[t!]
    \centering
    \includegraphics[width=0.91\linewidth]{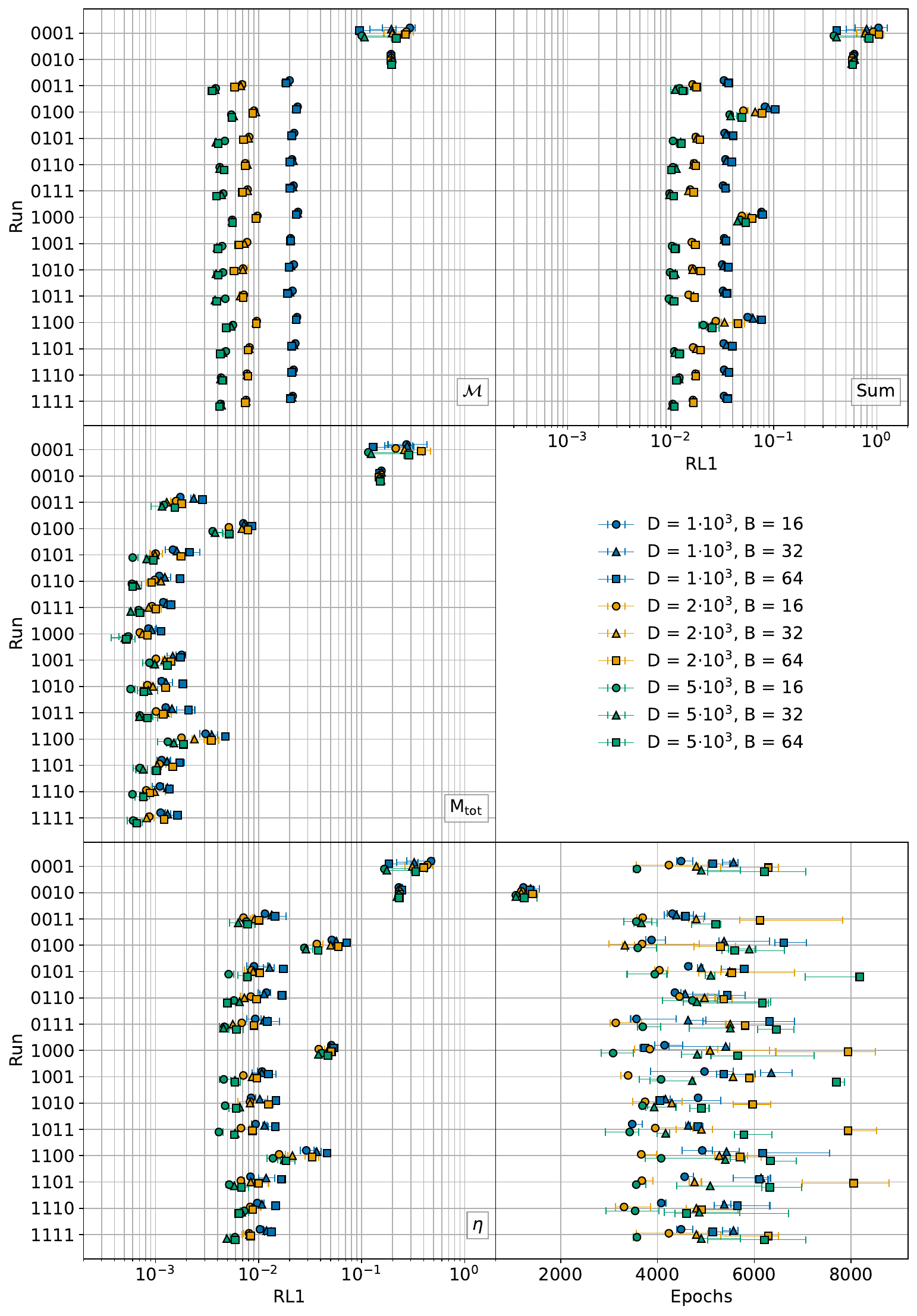}
    \caption{\emph{RL1 results (median and 68\% CI) for all the runs. In the $y$ axis the runs are labeled following their $\left\{\alpha_\mathrm{t},\alpha_\mathrm{df},\alpha_\mathrm{p},\alpha_\mathrm{a}\right\}$ values. The different panels present the study of the mass parameters individually, while the shape and the color of the markers determines the training dataset and the batch sizes. One can clearly notice that when APRIL losses are absent the RL1 result for the common sum is worse than a factor 10. Furthermore, there is a clear pattern: the higher $D$ and the lower $B$, the better the results. Notice also, that most of the effect due to the APRIL terms is related to $\eta$. It is this stiff parameter, hard to train otherwise, which is guided to better converge due to relations with other mass parameters implemented inside the redundant losses.}}
    \label{fig:all_runs}
\end{figure}

Our sampling is performed with the rejection sampling \cite{vonNeumann1951,RobertCasella2004}, a technique for generating samples from a distribution that is difficult to sample directly. One can always take a simpler distribution, sample from that and check if the sampled value is coherent with the target distribution. In practice, we sample $m_1^*$ from a uniform distribution $\mathcal{U}\left(\left[35.5,50.0\right]\text{ M}_\odot\right)$ and a value $u^*$ from $\mathcal{U}\left(\left[0,1\right]\right)$. Thanks to the $\pi\left(m_1\right)$ from Eq.~\ref{eq:p(m1)}, we can calculate the probability of having $m_1^*$, i.e. $\pi\left(m_1^*\right)$, and the maximum value of the probability in the $m_1$ domain, i.e. $\max\left[\pi\left(m_1\right)\right]$. If $u^*<\tfrac{\pi\left(m_1^*\right)}{\max\left[\pi\left(m_1\right)\right]}$, $m_1^*$ is kept as sample from $\pi\left(m_1\right)$ probability distribution. The same procedure is then done for $q$, using $\pi\left(q|m_1\right)$ in Eq.~\ref{eq:p(q)} in the range  $q\in[0.127,1.000]$. Since we used the rejection sampling, where a ratio between each probability and its maximum value was considered, no normalization constant was implemented for Eqs.~\ref{eq:p(m1)} and \ref{eq:p(q)}.

\section{Extended results}
\label{app:extended_results}
Here we present extended results which complete results in  Sec.~\ref{sec:discussion}. The total loss considered in this case is different from the \cref{eq:L_total_bench}, i.e.
\begin{equation}
    \mathcal{L}_\mathrm{total}(\theta)=\alpha_\mathrm{t}\mathcal{L}_\mathrm{t}(\theta)+\mathcal{L}_\mathrm{APRIL}(\theta)=\alpha_\mathrm{t}\mathcal{L}_\mathrm{t}(\theta)+\alpha_\mathrm{df}\mathcal{L}_\mathrm{df}(\theta)+\alpha_\mathrm{p}\mathcal{L}_\mathrm{p}(\theta)+\alpha_\mathrm{a}\mathcal{L}_\mathrm{a}(\theta)\,.
\end{equation}

Note how the coefficients are distributed among different terms. Switching on and off different hyper-parameters, we can benchmark all the different terms. The benchmark and test procedure is exactly the same as presented in Sec.~\ref{sec:Test_description}, with the only difference that this time we are performing 15 runs instead of 3 per $\left\{D,B\right\}$ couple. We are considering all possible $\alpha_\mathrm{i}$ combinations, except $\left\{\alpha_\mathrm{t},\alpha_\mathrm{df},\alpha_\mathrm{p},\alpha_\mathrm{a}\right\}=\left\{0,0,0,0\right\}$.

The results are shown in Fig.~\ref{fig:all_runs} and Table~\ref{tab:test_runs}. In particular, in the y-axis of Fig.~\ref{fig:all_runs} the runs are listed following $\left\{\alpha_\mathrm{t},\alpha_\mathrm{df},\alpha_\mathrm{p},\alpha_\mathrm{a}\right\}$, while the rank in Table~\ref{tab:test_runs} is performed both locally (for each $\left\{D,B\right\}$ couple) and globally. The presence of $\mathcal{L}_\mathrm{p}$ and $\mathcal{L}_\mathrm{a}$ is helping the performance by a factor 10 with respect to their absence. In particular, one can see the systematic of this approach: Sec.~\ref{sec:discussion} results are not only due to the particular loss combination, but are consistent even when switching on and off the various terms.

The influence of the $\mathcal{L}_\mathrm{df}$ term deserves an additional discussion. It still improves the results with respect to runs where it is absent (see for example 1000 and 1100 runs in Fig.~\ref{fig:all_runs}), but this is not so relevant as for $\mathcal{L}_\mathrm{p}$ and $\mathcal{L}_\mathrm{a}$. This could be given by the fact that in $\mathcal{L}_\mathrm{df}$ the relations between parameters are more complex and the difference in magnitude between its hidden terms ($\mathcal{L}_\textrm{Newt}$ and $\mathcal{L}_\textrm{corr}$) is hard to handle. This complexity could be, at the same time, the reason why $\mathcal{L}_\mathbf{df}$ alone performs better than 0001 and 0010 runs. The results are showing that it is hard to combine $\mathcal{L}_\mathbf{df}$ with other terms, but the complexity of the differential equation is sufficient to model the GW frequency behavior in a correct way. An interesting followup study will be a hyperparameter fine-tuning or active optimization during training for the two coefficients for the two hidden terms, $\beta$ and $\gamma$ in Eq.~\ref{eq:L_df}. Furthermore, the use of geometric units ($c=G=1$) will lead to more balanced magnitudes. Here we chose to use physical quantities in order to study GW signals as similar as possible to the detected events.

\begin{small}
    \begin{longtable}{c|rrrr|rr|r}
        \caption{\emph{RL1 performance on the test dataset for the different models. The values corresponds to the median and the $68\%$ CI and are plotted in Fig.~\ref{fig:all_runs}.}}
        \label{tab:test_runs} \\
        \multicolumn{1}{c|}{\emph{Run}}&\multicolumn{4}{c|}{\emph{RL1 ($\times10^{-3}$)}}&\multicolumn{2}{c|}{\emph{Rank}}&\multicolumn{1}{c}{\emph{Epochs}}\\
        \cmidrule{1-7}
        $\alpha_\mathrm{t}\alpha_\mathrm{df}\alpha_\mathrm{p}\alpha_\mathrm{a}$&\multicolumn{1}{c}{$\mathcal{M}$}&\multicolumn{1}{c}{$M_\mathrm{tot}$}&\multicolumn{1}{c}{$\eta$}&\multicolumn{1}{c|}{\emph{Sum}}&\multicolumn{1}{c}{\emph{L.}}&\multicolumn{1}{c|}{\emph{G.}}&\\
        \toprule
        \endfirsthead
        
        \multicolumn{1}{c|}{\emph{Run}}&\multicolumn{4}{c|}{\emph{RL1 ($\times10^{-3}$)}}&\multicolumn{2}{c|}{\emph{Rank}}&\multicolumn{1}{c}{\emph{Epochs}}\\
        \cmidrule{1-7}
        $\alpha_\mathrm{t}\alpha_\mathrm{df}\alpha_\mathrm{p}\alpha_\mathrm{a}$&\multicolumn{1}{c}{$\mathcal{M}$}&\multicolumn{1}{c}{$M_\mathrm{tot}$}&\multicolumn{1}{c}{$\eta$}&\multicolumn{1}{c|}{\emph{Sum}}&\multicolumn{1}{c}{\emph{L.}}&\multicolumn{1}{c|}{\emph{G.}}&\\
        \toprule
        \endhead

        \midrule
        \multicolumn{8}{r}{\emph{Continue in next page...}} \\
        \endfoot
        
        \bottomrule
        \endlastfoot
        
        \multicolumn{8}{c}{$D=10^3$, $B=16$} \\
        \midrule
        0\,\,\,\,0\,\,\,\,0\,\,\,\,1&$294.13^{+34.41}_{-135.93}$&$272.73^{+155.89}_{-92.21}$&$471.88^{+33.50}_{-196.09}$&$1038.75^{+223.81}_{-424.24}$&15&134&$3550^{+180}_{-307}$\\
0\,\,\,\,0\,\,\,\,1\,\,\,\,0&$193.37^{+3.17}_{-0.77}$&$155.75^{+1.10}_{-5.80}$&$229.61^{+19.97}_{-3.49}$&$602.82^{+5.72}_{-20.15}$&14&128&$1232^{+25}_{-19}$\\
0\,\,\,\,0\,\,\,\,1\,\,\,\,1&$19.99^{+0.04}_{-0.28}$&$1.74^{+0.06}_{-0.06}$&$11.54^{+0.31}_{-0.62}$&$32.81^{+0.55}_{-0.34}$&5&70&$4307^{+101}_{-104}$\\
0\,\,\,\,1\,\,\,\,0\,\,\,\,0&$23.93^{+0.50}_{-0.34}$&$7.12^{+0.23}_{-0.28}$&$51.31^{+3.18}_{-1.09}$&$81.98^{+3.03}_{-1.15}$&13&115&$3877^{+282}_{-114}$\\
0\,\,\,\,1\,\,\,\,0\,\,\,\,1&$22.13^{+0.41}_{-1.02}$&$1.48^{+0.07}_{-0.22}$&$9.03^{+3.29}_{-1.37}$&$33.30^{+1.80}_{-1.66}$&9&77&$4637^{+72}_{-30}$\\
0\,\,\,\,1\,\,\,\,1\,\,\,\,0&$21.14^{+1.02}_{-0.10}$&$1.09^{+0.04}_{-0.00}$&$11.90^{+1.13}_{-1.98}$&$33.95^{+1.06}_{-0.81}$&10&78&$4360^{+123}_{-8}$\\
0\,\,\,\,1\,\,\,\,1\,\,\,\,1&$21.84^{+0.29}_{-0.84}$&$1.19^{+0.10}_{-0.07}$&$9.29^{+1.29}_{-1.61}$&$32.12^{+1.45}_{-1.08}$&2&66&$3566^{+823}_{-120}$\\
1\,\,\,\,0\,\,\,\,0\,\,\,\,0&$24.14^{+0.00}_{-0.00}$&$0.85^{+0.17}_{-0.00}$&$50.90^{+2.40}_{-1.19}$&$75.89^{+2.20}_{-1.31}$&12&111&$4146^{+377}_{-198}$\\
1\,\,\,\,0\,\,\,\,0\,\,\,\,1&$20.43^{+0.00}_{-0.26}$&$1.79^{+0.17}_{-0.20}$&$10.82^{+0.48}_{-1.07}$&$33.04^{+0.53}_{-1.18}$&8&75&$4971^{+593}_{-1118}$\\
1\,\,\,\,0\,\,\,\,1\,\,\,\,0&$21.98^{+0.03}_{-0.70}$&$1.15^{+0.00}_{-0.08}$&$8.46^{+0.89}_{-0.50}$&$31.59^{+0.26}_{-0.49}$&1&65&$4838^{+41}_{-50}$\\
1\,\,\,\,0\,\,\,\,1\,\,\,\,1&$21.35^{+0.05}_{-0.00}$&$1.26^{+0.07}_{-0.00}$&$9.36^{+0.33}_{-0.79}$&$32.14^{+0.19}_{-1.10}$&3&67&$3478^{+210}_{-35}$\\
1\,\,\,\,1\,\,\,\,0\,\,\,\,0&$23.62^{+0.28}_{-0.14}$&$3.06^{+0.42}_{-0.37}$&$29.06^{+5.50}_{-3.64}$&$55.74^{+5.62}_{-3.31}$&11&105&$4924^{+212}_{-413}$\\
1\,\,\,\,1\,\,\,\,0\,\,\,\,1&$22.72^{+0.04}_{-0.35}$&$1.15^{+0.26}_{-0.11}$&$8.34^{+0.35}_{-0.17}$&$32.57^{+0.02}_{-0.44}$&4&69&$4561^{+179}_{-9}$\\
1\,\,\,\,1\,\,\,\,1\,\,\,\,0&$21.87^{+0.41}_{-0.14}$&$1.10^{+0.09}_{-0.18}$&$9.72^{+1.54}_{-0.33}$&$32.97^{+0.86}_{-0.29}$&7&74&$4078^{+93}_{-41}$\\
1\,\,\,\,1\,\,\,\,1\,\,\,\,1&$21.43^{+0.10}_{-0.69}$&$1.12^{+0.13}_{-0.04}$&$10.31^{+0.79}_{-0.50}$&$32.86^{+0.23}_{-0.19}$&6&72&$4488^{+240}_{-91}$\\
        \toprule
        
        \multicolumn{8}{c}{$D=10^3$, $B=32$} \\
        \midrule
        0\,\,\,\,0\,\,\,\,0\,\,\,\,1&$194.46^{+22.37}_{-74.15}$&$277.14^{+36.83}_{-109.72}$&$323.53^{+29.99}_{-104.44}$&$795.13^{+89.19}_{-288.32}$&15&131&$5645^{+240}_{-720}$\\
0\,\,\,\,0\,\,\,\,1\,\,\,\,0&$193.40^{+0.54}_{-1.61}$&$156.02^{+1.18}_{-7.51}$&$233.94^{+16.63}_{-0.61}$&$595.65^{+4.46}_{-12.07}$&14&126&$1382^{+176}_{-34}$\\
0\,\,\,\,0\,\,\,\,1\,\,\,\,1&$19.11^{+0.68}_{-0.14}$&$2.35^{+0.07}_{-0.14}$&$13.40^{+0.29}_{-0.00}$&$34.89^{+0.40}_{-0.03}$&10&87&$4391^{+352}_{-256}$\\
0\,\,\,\,1\,\,\,\,0\,\,\,\,0&$23.17^{+0.63}_{-0.14}$&$7.22^{+0.15}_{-0.48}$&$56.64^{+1.49}_{-3.67}$&$88.11^{+0.57}_{-4.65}$&13&116&$5375^{+941}_{-101}$\\
0\,\,\,\,1\,\,\,\,0\,\,\,\,1&$21.12^{+0.26}_{-0.11}$&$1.61^{+0.00}_{-0.02}$&$12.80^{+1.28}_{-1.62}$&$34.42^{+1.73}_{-0.81}$&6&83&$4902^{+39}_{-73}$\\
0\,\,\,\,1\,\,\,\,1\,\,\,\,0&$21.78^{+0.23}_{-0.42}$&$1.23^{+0.17}_{-0.09}$&$11.32^{+0.53}_{-0.98}$&$34.72^{+0.07}_{-1.30}$&8&85&$4573^{+696}_{-45}$\\
0\,\,\,\,1\,\,\,\,1\,\,\,\,1&$21.63^{+0.12}_{-1.03}$&$1.28^{+0.15}_{-0.06}$&$11.28^{+1.34}_{-1.72}$&$34.42^{+0.14}_{-1.63}$&5&82&$4633^{+365}_{-17}$\\
1\,\,\,\,0\,\,\,\,0\,\,\,\,0&$23.95^{+0.03}_{-0.08}$&$0.92^{+0.08}_{-0.11}$&$49.00^{+0.45}_{-1.46}$&$73.71^{+0.43}_{-1.42}$&12&110&$5410^{+78}_{-1171}$\\
1\,\,\,\,0\,\,\,\,0\,\,\,\,1&$20.64^{+0.85}_{-0.19}$&$1.47^{+0.00}_{-0.17}$&$10.75^{+0.45}_{-2.10}$&$32.85^{+0.00}_{-1.18}$&2&71&$6351^{+427}_{-224}$\\
1\,\,\,\,0\,\,\,\,1\,\,\,\,0&$20.66^{+0.19}_{-0.38}$&$1.26^{+0.19}_{-0.13}$&$10.29^{+1.97}_{-0.03}$&$32.47^{+1.64}_{-0.19}$&1&68&$4165^{+103}_{-33}$\\
1\,\,\,\,0\,\,\,\,1\,\,\,\,1&$20.17^{+0.12}_{-0.37}$&$1.45^{+0.00}_{-0.03}$&$11.41^{+0.60}_{-0.40}$&$32.90^{+0.10}_{-0.19}$&3&73&$4638^{+48}_{-53}$\\
1\,\,\,\,1\,\,\,\,0\,\,\,\,0&$23.24^{+0.70}_{-0.11}$&$3.52^{+0.50}_{-0.26}$&$36.66^{+3.96}_{-1.11}$&$62.91^{+5.54}_{-1.09}$&11&108&$5426^{+258}_{-444}$\\
1\,\,\,\,1\,\,\,\,0\,\,\,\,1&$22.12^{+0.26}_{-0.24}$&$1.30^{+0.06}_{-0.02}$&$11.85^{+2.58}_{-1.78}$&$34.84^{+2.78}_{-1.21}$&9&86&$6139^{+194}_{-578}$\\
1\,\,\,\,1\,\,\,\,1\,\,\,\,0&$21.44^{+0.69}_{-0.33}$&$1.30^{+0.10}_{-0.00}$&$10.79^{+0.22}_{-0.07}$&$34.39^{+0.29}_{-0.21}$&4&81&$5379^{+140}_{-32}$\\
1\,\,\,\,1\,\,\,\,1\,\,\,\,1&$21.16^{+0.07}_{-0.86}$&$1.31^{+0.09}_{-0.10}$&$11.96^{+1.30}_{-0.52}$&$34.54^{+0.35}_{-0.53}$&7&84&$5570^{+89}_{-231}$\\
        \toprule
        
        \multicolumn{8}{c}{$D=10^3$, $B=64$} \\
        \midrule
        0\,\,\,\,0\,\,\,\,0\,\,\,\,1&$95.46^{+120.18}_{-2.01}$&$129.08^{+192.34}_{-2.36}$&$183.79^{+164.84}_{-2.74}$&$408.33^{+477.36}_{-7.11}$&14&120&$3359^{+2112}_{-379}$\\
0\,\,\,\,0\,\,\,\,1\,\,\,\,0&$192.68^{+6.39}_{-1.82}$&$148.87^{+2.57}_{-3.25}$&$244.91^{+3.07}_{-7.68}$&$590.02^{+1.31}_{-7.51}$&15&125&$1364^{+102}_{-8}$\\
0\,\,\,\,0\,\,\,\,1\,\,\,\,1&$18.52^{+0.00}_{-1.66}$&$2.86^{+0.24}_{-0.16}$&$14.44^{+4.06}_{-0.19}$&$36.42^{+2.22}_{-0.77}$&6&91&$4579^{+392}_{-257}$\\
0\,\,\,\,1\,\,\,\,0\,\,\,\,0&$23.31^{+0.47}_{-1.03}$&$8.60^{+0.33}_{-0.59}$&$71.57^{+1.33}_{-5.39}$&$103.24^{+0.50}_{-5.30}$&13&117&$6607^{+469}_{-181}$\\
0\,\,\,\,1\,\,\,\,0\,\,\,\,1&$21.01^{+0.00}_{-1.00}$&$2.14^{+0.56}_{-0.06}$&$17.40^{+1.39}_{-1.24}$&$40.46^{+1.02}_{-1.10}$&10&97&$5794^{+40}_{-480}$\\
0\,\,\,\,1\,\,\,\,1\,\,\,\,0&$20.12^{+0.72}_{-0.15}$&$1.73^{+0.15}_{-0.15}$&$16.90^{+1.57}_{-1.64}$&$38.96^{+1.11}_{-0.84}$&8&95&$5440^{+366}_{-708}$\\
0\,\,\,\,1\,\,\,\,1\,\,\,\,1&$20.16^{+0.23}_{-0.72}$&$1.42^{+0.08}_{-0.15}$&$12.15^{+3.77}_{-0.51}$&$34.05^{+2.67}_{-0.12}$&1&79&$6312^{+513}_{-1374}$\\
1\,\,\,\,0\,\,\,\,0\,\,\,\,0&$23.16^{+0.19}_{-0.40}$&$1.13^{+0.00}_{-0.07}$&$53.71^{+0.10}_{-0.97}$&$78.14^{+0.00}_{-1.21}$&12&114&$3740^{+17}_{-107}$\\
1\,\,\,\,0\,\,\,\,0\,\,\,\,1&$20.49^{+0.32}_{-0.75}$&$1.75^{+0.00}_{-0.25}$&$12.49^{+2.27}_{-1.18}$&$34.38^{+1.77}_{-0.43}$&2&80&$5366^{+647}_{-150}$\\
1\,\,\,\,0\,\,\,\,1\,\,\,\,0&$19.71^{+0.03}_{-0.22}$&$1.84^{+0.18}_{-0.06}$&$14.78^{+0.32}_{-0.09}$&$36.20^{+0.82}_{-0.23}$&5&90&$4051^{+1253}_{-37}$\\
1\,\,\,\,0\,\,\,\,1\,\,\,\,1&$19.16^{+0.12}_{-0.30}$&$2.09^{+0.31}_{-0.50}$&$14.55^{+0.85}_{-2.18}$&$35.00^{+1.43}_{-1.58}$&3&88&$4836^{+103}_{-154}$\\
1\,\,\,\,1\,\,\,\,0\,\,\,\,0&$23.13^{+0.19}_{-0.00}$&$4.76^{+0.04}_{-0.30}$&$46.19^{+2.78}_{-1.95}$&$76.20^{+1.26}_{-3.79}$&11&112&$6171^{+1380}_{-8}$\\
1\,\,\,\,1\,\,\,\,0\,\,\,\,1&$20.95^{+0.14}_{-0.05}$&$1.71^{+0.18}_{-0.05}$&$16.68^{+0.50}_{-0.95}$&$39.67^{+0.26}_{-1.05}$&9&96&$6103^{+181}_{-85}$\\
1\,\,\,\,1\,\,\,\,1\,\,\,\,0&$20.97^{+0.13}_{-0.22}$&$1.37^{+0.10}_{-0.10}$&$14.68^{+0.94}_{-0.50}$&$36.88^{+0.67}_{-0.18}$&7&92&$5649^{+672}_{-477}$\\
1\,\,\,\,1\,\,\,\,1\,\,\,\,1&$20.45^{+0.87}_{-0.18}$&$1.63^{+0.12}_{-0.31}$&$13.37^{+0.61}_{-1.97}$&$35.62^{+0.00}_{-0.95}$&4&89&$5142^{+355}_{-80}$\\
        \toprule
        
        \multicolumn{8}{c}{$D=2\cdot10^3$, $B=16$} \\
        \midrule
        0\,\,\,\,0\,\,\,\,0\,\,\,\,1&$272.34^{+43.83}_{-76.00}$&$214.43^{+76.20}_{-3.80}$&$433.99^{+52.67}_{-119.25}$&$920.77^{+172.70}_{-199.05}$&15&133&$1974^{+2439}_{-215}$\\
0\,\,\,\,0\,\,\,\,1\,\,\,\,0&$194.96^{+2.91}_{-2.51}$&$153.98^{+1.62}_{-2.84}$&$230.50^{+19.31}_{-5.82}$&$578.19^{+20.99}_{-6.55}$&14&122&$1187^{+60}_{-58}$\\
0\,\,\,\,0\,\,\,\,1\,\,\,\,1&$6.91^{+0.62}_{-0.86}$&$1.59^{+0.07}_{-0.19}$&$7.14^{+1.52}_{-0.23}$&$16.28^{+0.19}_{-0.50}$&5&36&$3694^{+21}_{-119}$\\
0\,\,\,\,1\,\,\,\,0\,\,\,\,0&$9.03^{+0.14}_{-0.34}$&$5.15^{+0.40}_{-0.35}$&$36.76^{+5.29}_{-1.42}$&$50.65^{+5.25}_{-0.75}$&13&103&$3681^{+1189}_{-684}$\\
0\,\,\,\,1\,\,\,\,0\,\,\,\,1&$8.11^{+0.09}_{-0.29}$&$1.00^{+0.16}_{-0.12}$&$8.58^{+0.55}_{-1.32}$&$17.48^{+0.58}_{-1.09}$&10&53&$4040^{+175}_{-99}$\\
0\,\,\,\,1\,\,\,\,1\,\,\,\,0&$7.43^{+0.65}_{-0.11}$&$0.98^{+0.00}_{-0.00}$&$8.41^{+1.32}_{-1.72}$&$16.83^{+1.21}_{-1.03}$&8&45&$4454^{+580}_{-28}$\\
0\,\,\,\,1\,\,\,\,1\,\,\,\,1&$7.82^{+0.00}_{-0.23}$&$0.93^{+0.00}_{-0.02}$&$6.87^{+0.12}_{-0.21}$&$15.46^{+0.10}_{-0.26}$&2&33&$3138^{+535}_{-110}$\\
1\,\,\,\,0\,\,\,\,0\,\,\,\,0&$9.74^{+0.12}_{-0.05}$&$0.70^{+0.09}_{-0.04}$&$38.37^{+5.71}_{-1.40}$&$48.95^{+5.77}_{-1.50}$&12&101&$3842^{+1398}_{-312}$\\
1\,\,\,\,0\,\,\,\,0\,\,\,\,1&$7.75^{+0.02}_{-0.03}$&$1.01^{+0.02}_{-0.03}$&$7.12^{+0.19}_{-0.41}$&$15.99^{+0.08}_{-0.61}$&3&34&$3394^{+24}_{-148}$\\
1\,\,\,\,0\,\,\,\,1\,\,\,\,0&$7.07^{+0.79}_{-0.12}$&$0.83^{+0.07}_{-0.07}$&$8.26^{+0.75}_{-1.96}$&$16.17^{+0.11}_{-0.97}$&4&35&$3743^{+760}_{-253}$\\
1\,\,\,\,0\,\,\,\,1\,\,\,\,1&$7.20^{+0.14}_{-0.69}$&$1.01^{+0.04}_{-0.04}$&$6.75^{+1.49}_{-0.14}$&$14.91^{+0.76}_{-0.06}$&1&32&$3954^{+777}_{-18}$\\
1\,\,\,\,1\,\,\,\,0\,\,\,\,0&$9.60^{+0.00}_{-0.15}$&$1.79^{+0.04}_{-0.04}$&$15.84^{+2.38}_{-0.68}$&$27.33^{+1.84}_{-0.86}$&11&64&$3665^{+315}_{-9}$\\
1\,\,\,\,1\,\,\,\,0\,\,\,\,1&$8.16^{+0.15}_{-0.03}$&$1.10^{+0.17}_{-0.08}$&$6.74^{+0.53}_{-0.00}$&$16.46^{+0.11}_{-0.20}$&7&39&$3678^{+224}_{-40}$\\
1\,\,\,\,1\,\,\,\,1\,\,\,\,0&$7.71^{+0.41}_{-0.34}$&$0.81^{+0.08}_{-0.00}$&$8.30^{+2.22}_{-0.39}$&$17.38^{+1.43}_{-0.72}$&9&50&$3310^{+548}_{-176}$\\
1\,\,\,\,1\,\,\,\,1\,\,\,\,1&$7.57^{+0.19}_{-0.05}$&$0.87^{+0.00}_{-0.04}$&$8.09^{+0.09}_{-0.85}$&$16.41^{+0.16}_{-0.56}$&6&38&$4236^{+801}_{-667}$\\
        \toprule
        
        \multicolumn{8}{c}{$D=2\cdot10^3$, $B=32$} \\
        \midrule
        0\,\,\,\,0\,\,\,\,0\,\,\,\,1&$198.11^{+51.12}_{-33.74}$&$260.13^{+91.12}_{-52.08}$&$310.61^{+68.88}_{-46.50}$&$768.85^{+211.11}_{-132.32}$&15&130&$6048^{+29}_{-343}$\\
0\,\,\,\,0\,\,\,\,1\,\,\,\,0&$195.01^{+3.77}_{-0.15}$&$150.68^{+6.44}_{-0.33}$&$228.66^{+20.99}_{-3.75}$&$609.25^{+2.30}_{-26.29}$&14&129&$1186^{+35}_{-42}$\\
0\,\,\,\,0\,\,\,\,1\,\,\,\,1&$6.73^{+0.08}_{-0.24}$&$1.29^{+0.15}_{-0.03}$&$9.14^{+0.21}_{-0.79}$&$16.83^{+0.30}_{-0.51}$&6&46&$4801^{+145}_{-160}$\\
0\,\,\,\,1\,\,\,\,0\,\,\,\,0&$9.41^{+0.19}_{-0.33}$&$6.88^{+0.71}_{-0.00}$&$50.14^{+4.72}_{-0.53}$&$65.98^{+6.36}_{-0.56}$&13&109&$3327^{+198}_{-48}$\\
0\,\,\,\,1\,\,\,\,0\,\,\,\,1&$8.02^{+0.11}_{-0.27}$&$0.99^{+0.09}_{-0.08}$&$8.87^{+0.22}_{-0.22}$&$17.84^{+0.14}_{-0.14}$&10&57&$5492^{+48}_{-180}$\\
0\,\,\,\,1\,\,\,\,1\,\,\,\,0&$7.80^{+0.05}_{-0.00}$&$1.13^{+0.06}_{-0.07}$&$7.29^{+1.17}_{-0.12}$&$16.53^{+0.87}_{-0.10}$&4&41&$4972^{+218}_{-474}$\\
0\,\,\,\,1\,\,\,\,1\,\,\,\,1&$7.85^{+0.10}_{-0.00}$&$0.86^{+0.03}_{-0.02}$&$5.62^{+1.36}_{-0.07}$&$14.83^{+0.36}_{-0.37}$&1&31&$5503^{+11}_{-91}$\\
1\,\,\,\,0\,\,\,\,0\,\,\,\,0&$9.44^{+0.03}_{-0.04}$&$0.75^{+0.03}_{-0.00}$&$48.14^{+0.93}_{-0.86}$&$57.93^{+1.41}_{-0.65}$&12&106&$5086^{+1228}_{-73}$\\
1\,\,\,\,0\,\,\,\,0\,\,\,\,1&$7.29^{+0.16}_{-0.36}$&$1.23^{+0.07}_{-0.06}$&$8.66^{+0.40}_{-1.59}$&$17.22^{+0.04}_{-1.40}$&7&48&$5561^{+32}_{-6}$\\
1\,\,\,\,0\,\,\,\,1\,\,\,\,0&$7.05^{+0.39}_{-0.18}$&$0.94^{+0.05}_{-0.09}$&$8.28^{+0.53}_{-0.34}$&$16.32^{+0.51}_{-0.22}$&2&37&$4297^{+4}_{-53}$\\
1\,\,\,\,0\,\,\,\,1\,\,\,\,1&$6.62^{+0.12}_{-0.12}$&$1.24^{+0.18}_{-0.14}$&$8.79^{+0.14}_{-0.63}$&$16.65^{+0.07}_{-0.00}$&5&43&$4908^{+225}_{-518}$\\
1\,\,\,\,1\,\,\,\,0\,\,\,\,0&$9.44^{+0.04}_{-0.13}$&$2.38^{+0.59}_{-0.12}$&$21.32^{+6.80}_{-2.10}$&$33.20^{+7.46}_{-2.27}$&11&76&$5279^{+584}_{-94}$\\
1\,\,\,\,1\,\,\,\,0\,\,\,\,1&$8.09^{+0.03}_{-0.29}$&$1.04^{+0.10}_{-0.02}$&$8.39^{+0.64}_{-1.08}$&$17.70^{+0.55}_{-1.22}$&9&55&$4765^{+143}_{-111}$\\
1\,\,\,\,1\,\,\,\,1\,\,\,\,0&$7.74^{+0.16}_{-0.05}$&$0.98^{+0.26}_{-0.05}$&$8.70^{+0.48}_{-0.23}$&$17.35^{+0.44}_{-0.29}$&8&49&$4803^{+189}_{-210}$\\
1\,\,\,\,1\,\,\,\,1\,\,\,\,1&$7.59^{+0.47}_{-0.10}$&$0.82^{+0.15}_{-0.00}$&$8.13^{+0.14}_{-1.19}$&$16.53^{+0.05}_{-0.40}$&3&40&$4803^{+40}_{-574}$\\
        \toprule
        
        \multicolumn{8}{c}{$D=2\cdot10^3$, $B=64$} \\
        \midrule
        0\,\,\,\,0\,\,\,\,0\,\,\,\,1&$266.19^{+11.74}_{-12.60}$&$380.78^{+86.43}_{-26.14}$&$401.52^{+15.43}_{-14.86}$&$1048.50^{+113.59}_{-53.60}$&15&135&$5937^{+400}_{-1264}$\\
0\,\,\,\,0\,\,\,\,1\,\,\,\,0&$191.75^{+2.08}_{-0.04}$&$148.25^{+5.73}_{-1.30}$&$232.29^{+16.55}_{-6.49}$&$580.25^{+17.28}_{-11.44}$&14&123&$1422^{+101}_{-145}$\\
0\,\,\,\,0\,\,\,\,1\,\,\,\,1&$5.84^{+0.28}_{-0.24}$&$1.80^{+0.03}_{-0.00}$&$10.11^{+0.02}_{-0.21}$&$17.78^{+0.24}_{-0.00}$&7&56&$6117^{+1707}_{-417}$\\
0\,\,\,\,1\,\,\,\,0\,\,\,\,0&$8.79^{+0.37}_{-0.30}$&$7.90^{+0.13}_{-0.03}$&$59.52^{+0.88}_{-1.09}$&$76.90^{+0.07}_{-1.26}$&13&113&$5301^{+316}_{-542}$\\
0\,\,\,\,1\,\,\,\,0\,\,\,\,1&$7.13^{+0.14}_{-0.02}$&$1.77^{+0.13}_{-0.16}$&$10.27^{+0.68}_{-0.88}$&$19.17^{+0.72}_{-0.67}$&8&58&$5535^{+1292}_{-687}$\\
0\,\,\,\,1\,\,\,\,1\,\,\,\,0&$7.42^{+0.07}_{-0.23}$&$0.92^{+0.04}_{-0.10}$&$9.58^{+0.06}_{-0.33}$&$17.53^{+0.39}_{-0.05}$&6&54&$5368^{+176}_{-134}$\\
0\,\,\,\,1\,\,\,\,1\,\,\,\,1&$6.98^{+0.14}_{-0.08}$&$1.01^{+0.13}_{-0.02}$&$9.03^{+0.14}_{-0.22}$&$16.70^{+0.50}_{-0.10}$&2&44&$5811^{+577}_{-69}$\\
1\,\,\,\,0\,\,\,\,0\,\,\,\,0&$9.44^{+0.18}_{-0.39}$&$0.83^{+0.02}_{-0.02}$&$50.89^{+0.29}_{-4.18}$&$61.51^{+0.00}_{-4.94}$&12&107&$7933^{+569}_{-1490}$\\
1\,\,\,\,0\,\,\,\,0\,\,\,\,1&$6.45^{+0.12}_{-0.04}$&$1.41^{+0.05}_{-0.21}$&$9.64^{+0.12}_{-0.17}$&$17.38^{+0.56}_{-0.08}$&4&51&$5896^{+101}_{-11}$\\
1\,\,\,\,0\,\,\,\,1\,\,\,\,0&$5.77^{+0.06}_{-0.02}$&$1.25^{+0.00}_{-0.05}$&$12.52^{+1.52}_{-0.74}$&$19.56^{+1.17}_{-0.56}$&10&60&$5967^{+369}_{-405}$\\
1\,\,\,\,0\,\,\,\,1\,\,\,\,1&$7.08^{+0.30}_{-0.11}$&$1.19^{+0.24}_{-0.10}$&$8.75^{+0.24}_{-0.31}$&$17.02^{+0.77}_{-0.66}$&3&47&$7934^{+587}_{-14}$\\
1\,\,\,\,1\,\,\,\,0\,\,\,\,0&$9.51^{+0.28}_{-0.58}$&$3.48^{+0.66}_{-0.37}$&$33.19^{+5.24}_{-2.63}$&$44.84^{+7.16}_{-1.67}$&11&99&$5706^{+431}_{-442}$\\
1\,\,\,\,1\,\,\,\,0\,\,\,\,1&$7.89^{+0.04}_{-0.47}$&$1.47^{+0.08}_{-0.08}$&$10.00^{+2.50}_{-0.46}$&$19.41^{+1.91}_{-0.12}$&9&59&$8052^{+733}_{-1068}$\\
1\,\,\,\,1\,\,\,\,1\,\,\,\,0&$7.91^{+0.27}_{-0.25}$&$0.88^{+0.08}_{-0.00}$&$8.76^{+0.08}_{-1.43}$&$17.47^{+0.04}_{-1.01}$&5&52&$4911^{+757}_{-1}$\\
1\,\,\,\,1\,\,\,\,1\,\,\,\,1&$7.44^{+0.14}_{-0.44}$&$1.21^{+0.00}_{-0.03}$&$8.40^{+0.23}_{-0.60}$&$16.58^{+1.02}_{-0.14}$&1&42&$6292^{+210}_{-981}$\\
        \toprule
        
        \multicolumn{8}{c}{$D=5\cdot10^3$, $B=16$} \\
        \midrule
        0\,\,\,\,0\,\,\,\,0\,\,\,\,1&$100.49^{+189.36}_{-2.81}$&$116.08^{+277.76}_{-3.33}$&$166.38^{+261.56}_{-4.00}$&$382.95^{+728.68}_{-10.15}$&14&118&$5316^{+346}_{-2198}$\\
0\,\,\,\,0\,\,\,\,1\,\,\,\,0&$198.16^{+1.57}_{-2.62}$&$151.07^{+5.80}_{-0.36}$&$229.20^{+13.32}_{-6.11}$&$596.93^{+6.79}_{-12.62}$&15&127&$1075^{+23}_{-10}$\\
0\,\,\,\,0\,\,\,\,1\,\,\,\,1&$3.84^{+0.09}_{-0.17}$&$1.21^{+0.29}_{-0.11}$&$7.61^{+0.06}_{-1.22}$&$12.16^{+0.65}_{-0.87}$&10&26&$3563^{+323}_{-258}$\\
0\,\,\,\,1\,\,\,\,0\,\,\,\,0&$5.44^{+0.18}_{-0.07}$&$3.59^{+0.11}_{-0.00}$&$27.83^{+1.32}_{-1.03}$&$37.28^{+1.13}_{-1.16}$&12&93&$3593^{+383}_{-216}$\\
0\,\,\,\,1\,\,\,\,0\,\,\,\,1&$4.71^{+0.11}_{-0.17}$&$0.60^{+0.08}_{-0.00}$&$5.12^{+0.63}_{-0.05}$&$10.51^{+0.47}_{-0.14}$&7&11&$3944^{+251}_{-572}$\\
0\,\,\,\,1\,\,\,\,1\,\,\,\,0&$4.20^{+0.28}_{-0.03}$&$0.59^{+0.04}_{-0.00}$&$5.76^{+0.26}_{-0.78}$&$10.50^{+0.34}_{-0.40}$&6&10&$4720^{+525}_{-615}$\\
0\,\,\,\,1\,\,\,\,1\,\,\,\,1&$4.54^{+0.16}_{-0.10}$&$0.69^{+0.00}_{-0.04}$&$4.66^{+0.05}_{-0.51}$&$9.72^{+0.15}_{-0.16}$&2&2&$3696^{+376}_{-109}$\\
1\,\,\,\,0\,\,\,\,0\,\,\,\,0&$5.55^{+0.15}_{-0.03}$&$0.55^{+0.02}_{-0.10}$&$41.06^{+2.97}_{-2.77}$&$47.17^{+2.90}_{-2.75}$&13&100&$3084^{+416}_{-244}$\\
1\,\,\,\,0\,\,\,\,0\,\,\,\,1&$4.43^{+0.17}_{-0.25}$&$0.87^{+0.16}_{-0.12}$&$4.60^{+1.23}_{-0.41}$&$10.30^{+0.95}_{-0.64}$&4&7&$4074^{+0}_{-452}$\\
1\,\,\,\,0\,\,\,\,1\,\,\,\,0&$4.52^{+0.08}_{-0.06}$&$0.57^{+0.06}_{-0.00}$&$4.74^{+0.20}_{-0.23}$&$9.83^{+0.20}_{-0.20}$&3&4&$3691^{+103}_{-22}$\\
1\,\,\,\,0\,\,\,\,1\,\,\,\,1&$4.76^{+0.11}_{-0.02}$&$0.70^{+0.21}_{-0.02}$&$4.12^{+0.42}_{-0.20}$&$9.58^{+0.36}_{-0.11}$&1&1&$3426^{+184}_{-506}$\\
1\,\,\,\,1\,\,\,\,0\,\,\,\,0&$5.64^{+0.03}_{-0.03}$&$1.32^{+0.26}_{-0.26}$&$13.77^{+1.52}_{-1.73}$&$20.70^{+1.83}_{-1.88}$&11&61&$4076^{+1427}_{-333}$\\
1\,\,\,\,1\,\,\,\,0\,\,\,\,1&$4.82^{+0.03}_{-0.18}$&$0.70^{+0.02}_{-0.10}$&$5.20^{+0.60}_{-0.39}$&$10.80^{+0.22}_{-0.16}$&8&15&$3562^{+192}_{-31}$\\
1\,\,\,\,1\,\,\,\,1\,\,\,\,0&$4.33^{+0.04}_{-0.03}$&$0.60^{+0.03}_{-0.00}$&$7.26^{+0.16}_{-0.64}$&$12.11^{+0.07}_{-0.59}$&9&25&$3539^{+492}_{-606}$\\
1\,\,\,\,1\,\,\,\,1\,\,\,\,1&$4.29^{+0.30}_{-0.20}$&$0.61^{+0.02}_{-0.08}$&$5.90^{+0.22}_{-0.57}$&$10.45^{+0.17}_{-0.12}$&5&9&$3576^{+26}_{-71}$\\
        \toprule
        
        \multicolumn{8}{c}{$D=5\cdot10^3$, $B=32$} \\
        \midrule
        0\,\,\,\,0\,\,\,\,0\,\,\,\,1&$106.62^{+88.44}_{-3.56}$&$123.72^{+131.67}_{-4.17}$&$175.83^{+125.78}_{-5.17}$&$406.17^{+345.89}_{-12.91}$&14&119&$4149^{+109}_{-140}$\\
0\,\,\,\,0\,\,\,\,1\,\,\,\,0&$195.33^{+0.81}_{-2.47}$&$154.36^{+0.45}_{-2.47}$&$220.42^{+2.69}_{-4.53}$&$570.41^{+17.54}_{-6.57}$&15&121&$1076^{+107}_{-30}$\\
0\,\,\,\,0\,\,\,\,1\,\,\,\,1&$3.80^{+0.22}_{-0.06}$&$1.16^{+0.05}_{-0.25}$&$6.35^{+0.15}_{-1.14}$&$11.01^{+0.41}_{-0.96}$&6&19&$3669^{+326}_{-157}$\\
0\,\,\,\,1\,\,\,\,0\,\,\,\,0&$5.74^{+0.00}_{-0.00}$&$3.78^{+0.65}_{-0.18}$&$28.83^{+4.77}_{-1.90}$&$38.11^{+5.60}_{-1.91}$&12&94&$5895^{+130}_{-428}$\\
0\,\,\,\,1\,\,\,\,0\,\,\,\,1&$3.86^{+0.29}_{-0.08}$&$0.82^{+0.04}_{-0.03}$&$7.81^{+0.74}_{-1.48}$&$12.49^{+0.76}_{-1.21}$&10&28&$5102^{+82}_{-113}$\\
0\,\,\,\,1\,\,\,\,1\,\,\,\,0&$4.20^{+0.15}_{-0.13}$&$0.66^{+0.08}_{-0.02}$&$6.54^{+0.12}_{-0.39}$&$11.32^{+0.34}_{-0.00}$&8&22&$4816^{+1524}_{-286}$\\
0\,\,\,\,1\,\,\,\,1\,\,\,\,1&$4.38^{+0.04}_{-0.11}$&$0.58^{+0.05}_{-0.02}$&$4.57^{+0.50}_{-0.03}$&$9.72^{+0.15}_{-0.05}$&1&3&$5505^{+51}_{-850}$\\
1\,\,\,\,0\,\,\,\,0\,\,\,\,0&$5.53^{+0.17}_{-0.00}$&$0.50^{+0.07}_{-0.13}$&$38.09^{+10.70}_{-0.11}$&$44.37^{+10.39}_{-0.41}$&13&98&$4820^{+89}_{-318}$\\
1\,\,\,\,0\,\,\,\,0\,\,\,\,1&$3.95^{+0.46}_{-0.03}$&$0.98^{+0.06}_{-0.10}$&$5.92^{+0.13}_{-0.53}$&$10.85^{+0.10}_{-0.19}$&4&16&$4718^{+35}_{-870}$\\
1\,\,\,\,0\,\,\,\,1\,\,\,\,0&$3.87^{+0.02}_{-0.20}$&$0.85^{+0.20}_{-0.18}$&$6.57^{+0.10}_{-0.04}$&$11.07^{+0.53}_{-0.26}$&7&21&$3932^{+448}_{-174}$\\
1\,\,\,\,0\,\,\,\,1\,\,\,\,1&$3.78^{+0.16}_{-0.07}$&$0.70^{+0.24}_{-0.05}$&$5.93^{+0.33}_{-0.13}$&$10.34^{+0.62}_{-0.02}$&3&8&$4168^{+85}_{-175}$\\
1\,\,\,\,1\,\,\,\,0\,\,\,\,0&$5.42^{+0.02}_{-0.06}$&$1.51^{+0.07}_{-0.22}$&$17.79^{+0.09}_{-2.52}$&$24.58^{+0.21}_{-2.58}$&11&62&$5401^{+403}_{-26}$\\
1\,\,\,\,1\,\,\,\,0\,\,\,\,1&$4.51^{+0.26}_{-0.08}$&$0.76^{+0.08}_{-0.11}$&$5.82^{+1.26}_{-0.68}$&$10.94^{+1.29}_{-0.08}$&5&18&$5090^{+1068}_{-691}$\\
1\,\,\,\,1\,\,\,\,1\,\,\,\,0&$4.27^{+0.10}_{-0.11}$&$0.76^{+0.11}_{-0.05}$&$6.86^{+1.13}_{-0.38}$&$12.03^{+1.03}_{-0.70}$&9&24&$4862^{+833}_{-725}$\\
1\,\,\,\,1\,\,\,\,1\,\,\,\,1&$4.38^{+0.10}_{-0.18}$&$0.66^{+0.08}_{-0.00}$&$4.93^{+1.01}_{-0.07}$&$10.25^{+0.56}_{-0.28}$&2&6&$4905^{+809}_{-19}$\\
        \toprule
        
        \multicolumn{8}{c}{$D=5\cdot10^3$, $B=64$} \\
        \midrule
        0\,\,\,\,0\,\,\,\,0\,\,\,\,1&$217.60^{+1.74}_{-19.52}$&$286.55^{+4.00}_{-31.38}$&$334.67^{+1.94}_{-27.39}$&$838.82^{+7.67}_{-78.30}$&15&132&$9457^{+391}_{-1931}$\\
0\,\,\,\,0\,\,\,\,1\,\,\,\,0&$195.19^{+4.30}_{-0.24}$&$151.85^{+4.14}_{-2.29}$&$231.57^{+3.38}_{-8.87}$&$583.31^{+14.46}_{-1.61}$&14&124&$1246^{+267}_{-24}$\\
0\,\,\,\,0\,\,\,\,1\,\,\,\,1&$3.54^{+0.04}_{-0.04}$&$1.54^{+0.10}_{-0.00}$&$7.83^{+1.47}_{-0.00}$&$13.10^{+1.56}_{-0.14}$&10&30&$5203^{+107}_{-493}$\\
0\,\,\,\,1\,\,\,\,0\,\,\,\,0&$5.51^{+0.21}_{-0.07}$&$5.19^{+0.05}_{-0.66}$&$37.81^{+3.15}_{-2.29}$&$48.97^{+2.44}_{-3.06}$&12&102&$5596^{+1022}_{-267}$\\
0\,\,\,\,1\,\,\,\,0\,\,\,\,1&$4.06^{+0.13}_{-0.11}$&$0.95^{+0.05}_{-0.14}$&$7.80^{+0.00}_{-0.14}$&$12.68^{+0.22}_{-0.18}$&9&29&$8178^{+60}_{-1129}$\\
0\,\,\,\,1\,\,\,\,1\,\,\,\,0&$4.65^{+0.10}_{-0.05}$&$0.60^{+0.00}_{-0.00}$&$4.98^{+0.34}_{-0.39}$&$10.13^{+0.41}_{-0.05}$&1&5&$6169^{+119}_{-1277}$\\
0\,\,\,\,1\,\,\,\,1\,\,\,\,1&$3.91^{+0.31}_{-0.12}$&$0.71^{+0.03}_{-0.00}$&$6.10^{+0.96}_{-0.92}$&$10.69^{+0.85}_{-0.57}$&3&13&$6454^{+364}_{-384}$\\
1\,\,\,\,0\,\,\,\,0\,\,\,\,0&$5.59^{+0.02}_{-0.03}$&$0.52^{+0.11}_{-0.03}$&$47.00^{+1.35}_{-1.38}$&$53.29^{+1.58}_{-1.65}$&13&104&$5657^{+1576}_{-552}$\\
1\,\,\,\,0\,\,\,\,0\,\,\,\,1&$4.05^{+0.15}_{-0.11}$&$1.30^{+0.09}_{-0.30}$&$5.89^{+0.73}_{-0.16}$&$11.02^{+1.13}_{-0.26}$&6&20&$7694^{+173}_{-35}$\\
1\,\,\,\,0\,\,\,\,1\,\,\,\,0&$4.06^{+0.17}_{-0.10}$&$0.77^{+0.00}_{-0.07}$&$6.05^{+0.66}_{-0.90}$&$10.66^{+0.73}_{-0.54}$&2&12&$4908^{+160}_{-236}$\\
1\,\,\,\,0\,\,\,\,1\,\,\,\,1&$3.94^{+0.15}_{-0.14}$&$0.84^{+0.21}_{-0.06}$&$5.81^{+0.29}_{-0.22}$&$10.78^{+0.21}_{-0.49}$&4&14&$5786^{+582}_{-194}$\\
1\,\,\,\,1\,\,\,\,0\,\,\,\,0&$4.84^{+0.47}_{-0.11}$&$1.87^{+0.16}_{-0.41}$&$18.48^{+4.05}_{-2.02}$&$25.20^{+4.10}_{-1.54}$&11&63&$6332^{+539}_{-28}$\\
1\,\,\,\,1\,\,\,\,0\,\,\,\,1&$4.26^{+0.11}_{-0.09}$&$1.02^{+0.03}_{-0.10}$&$6.83^{+1.18}_{-0.00}$&$12.16^{+1.18}_{-0.02}$&8&27&$6322^{+657}_{-165}$\\
1\,\,\,\,1\,\,\,\,1\,\,\,\,0&$4.47^{+0.05}_{-0.11}$&$0.76^{+0.02}_{-0.13}$&$6.45^{+0.66}_{-0.48}$&$11.33^{+0.99}_{-0.36}$&7&23&$4603^{+2105}_{-240}$\\
1\,\,\,\,1\,\,\,\,1\,\,\,\,1&$4.17^{+0.24}_{-0.16}$&$0.66^{+0.05}_{-0.00}$&$5.94^{+0.52}_{-0.48}$&$10.86^{+0.35}_{-0.31}$&5&17&$6210^{+851}_{-1175}$\\
    \end{longtable}
\end{small}

\addcontentsline{toc}{section}{References}
\bibliographystyle{ieeetr}
\bibliography{ref}

\end{document}